\documentclass[12pt]{iopart}

\usepackage{iopams,color,epsfig}  
\usepackage[all]{xypic}

\newcommand{\be}{ \begin{equation}}
\newcommand{\ee}{\end{equation}}
\newcommand{\bea}[1]{\begin{eqnarray}\label{#1} }
\newcommand{\eea}{\end{eqnarray}}

\newcommand{\hs}[1]{\mbox{hs$[#1]$}}
\newcommand{\w}[1]{\mbox{$\mathcal{W}_\infty[#1]$}}
\newcommand{\W}{{\cal W}}
\newcommand{\un}{\mathfrak{u}}

\begin{document}

\title{Minimal Model Holography}

\author{Matthias R. Gaberdiel$^1$, Rajesh Gopakumar$^2$}

\address{$^1$Institut f\"ur Theoretische Physik, ETH Zurich, 
$\;$CH-8093 Z\"urich, Switzerland}
\address{$^2$Harish-Chandra Research Institute, 
$\;$Chhatnag Road, Jhusi,
$\;$Allahabad, India 211019}
\ead{gaberdiel@itp.phys.ethz.ch, gopakumr@hri.res.in}
\begin{abstract}
We review the duality relating 2d $\W_N$ minimal model CFTs, in a large $N$ 't~Hooft like
limit, to higher spin gravitational theories on AdS$_3$.  
\end{abstract}

\maketitle

\section{Introduction}

The search for simple examples of holography is important in the effort to penetrate the 
AdS/CFT correspondence. It involves seeking a hard-to-achieve balance between analytic 
tractability and intrinsic complexity. One wants to be able to capture enough of the physics of 
holography, especially of the aspects relevant to the puzzles of quantum gravity, with 
quantitive precision so as to be able to transfer the resulting understanding to more 
`realistic examples'. 

In this article, we review one such attempt in this search which appears to have a number of 
promising features. It is a particular instance of the general class of examples involving 
Vasiliev higher spin gauge theories  on AdS  with dual vector-like CFTs (in a large $N$ limit). 
The articles in this issue discuss various aspects as well as examples of higher spin holography. 
Here we focus on the specific case of a class of interacting vector like 2d (generically 
non-supersymmetric) CFTs and their AdS$_3$ duals in terms of a higher spin gauge theory 
coupled to matter fields.  

Two dimensional CFTs are among the best understood nontrivial quantum field theories 
\cite{Belavin:1984vu} and, moreover, have wide applications in diverse areas of physics. 
Since one has a high degree of analytic control over these theories, they can potentially 
provide a rich source of CFTs with interesting bulk AdS$_3$ duals. Of course, an essential 
ingredient in having a classical bulk dual is to have a large number of degrees of freedom 
as in a large $N$ vector or matrix theory. It is in such a family of theories that one can recover 
classical gravitational physics (not necessarily described by an Einstein Lagrangian) in a 
parametrically controlled manner from the finite $N$ quantum regime. 

However, systematic studies of the large $N$ limit of families of 2d CFTs have not been 
carried out until recently. One can imagine at least two categories of such theories: these  
are the vector-like, and the gauge-like models whose number of degrees of freedom 
(i.e.\ the central charge) scales as $N$ or $N^2$, respectively; here $N$ is the rank of 
some underlying gauge group. In complexity the former are obviously simpler, as is
familiar from the usual large $N$ vector models.
Nevertheless, even these are quite intricate in their detailed structure as we will see in this 
article. Thus these theories may strike a good balance between complexity and tractability. 
We will only briefly mention the case of the matrix-like families, which have just begun to be 
analysed, see e.g.\ \cite{Gopakumar:2012gd}, at the end of this review.
\smallskip

%\noindent 

More specifically, the family of theories we will be considering are so-called coset CFTs of the 
form
\begin{equation}\label{ourcos}
\frac{{{\rm SU}(N)}_k \otimes {\rm SU}(N)_1 }{ {\rm SU}(N)_{k+1}} \ .
\end{equation}
They have central charge 
\be\label{coscent}
c_{N,k}= (N-1)\Bigl[1- \frac{N(N+1)}{(N+k)(N+k+1)}\Bigr] \leq  (N-1) \ ,
\ee
and hence are vector-like. We will review many of the already known properties of these
CFTs in Sec.~2.2.  In our context the most important characteristic is that they have conserved
higher spin currents $W^{s}(z)$ with $s=3\ldots N$; their symmetry algebra is therefore
a ${\cal W}$-algebra, and the models (\ref{ourcos}) are usually referred to as the 
$\W_N$ minimal models. It is an important feature of 2d quantum field theories (and CFTs in particular)
that higher spin conserved currents are compatible with interactions ---  this is for example
not the case in $3$d \cite{Maldacena:2011jn}. We will review some of the salient facts 
about the ${\cal W}_N$ algebras in Sec.~3.1; as we will explain there, these algebras
are all special cases 
of an extended symmetry algebra known as $\w{\mu}$ which typically has all 
integer spins $s\geq 2$, and which can be truncated to $\W_N$ for $\mu=N$.

We will be interested, as mentioned, in the large $N$ limit of these theories. We shall
consider a 't~Hooft like limit, where we take $N,k\rightarrow \infty$ while 
keeping the 't~Hooft coupling 
\be\label{thftdef}
0 \leq \lambda=\frac{N}{N+k} \leq 1
\ee
fixed. Note that in this limit the central charge in (\ref{coscent}) behaves as $c =N(1-\lambda^2)$. 
We will describe, as we go along, some of the evidence that this limit is well behaved;
for instance, in Sec.~5.2, 5.3, 5.4 we will study the spectrum of operators in this limit, while 
in Sec.~6.1 we will review some of the results from studies of correlation functions. We will see 
that an appropriate part of the spectrum will organise itself, at large $N$, into a Fock space of 
multiparticle states. The correlation functions, in turn, will exhibit, rather nontrivially, the 
factorisation required for a good large $N$ limit.
\smallskip

Let us now turn to the bulk AdS theories  that are believed to be dual to these
minimal models. They are gravitational theories in  
AdS$_3$, containing one additional higher spin $s>2$ gauge field (for each $s$) 
together with some 
scalar fields. Theories of this kind were constructed by Vasiliev first in AdS$_4$
\cite{Vasiliev:1989re}, 
and then generalised to other dimensions including AdS$_3$
\cite{Vasiliev:1995dn,Vasiliev:1999ba}. In 3d, they are labelled 
by a single parameter $\mu$ and based on a higher spin gauge group known as $\hs{\mu}$
\cite{Prokushkin:1998bq,Prokushkin:1998vn};
we summarise some of the relevant facts about these theories and their symmetries in 
Sec.~2.1. As is familiar from the classic calculation of Brown \& Henneaux \cite{Brown:1986nw},
partial information about the dual CFT comes from the analysis of the asymptotic
symmetry algebra. For the case of the $\hs{\mu}$ theory, this symmetry algebra
was determined in \cite{Henneaux:2010xg, Campoleoni:2010zq, Gaberdiel:2011wb} 
and shown to define a classical Poisson algebra which agrees, in the classical
($c\rightarrow \infty$) limit, with $\w{\mu}$; this will be reviewed in Sec.~3.2. 

Based on this observation, it was proposed in \cite{Gaberdiel:2010pz} that the
$\hs{\mu}$ higher spin theory in AdS$_3$ is dual to the above 
't~Hooft limit of the $\W_N$ minimal models, where the 't~Hooft coupling $\lambda$
agrees with $\mu=\lambda$. Furthermore, in order to account for the full spectrum
of the minimal model CFTs, it was proposed that the higher spin theory is coupled to 
{\it two} complex scalar fields. Unlike the higher dimensional case, the scalar field is, in 3d,
not part of the higher spin multiplet, and hence does not need to be included from the start. 
However, in order to couple it consistently to the higher spin theory based on
$\hs{\mu}$, its mass is fixed to equal
$M^2=-1+\mu^2$  \cite{Prokushkin:1998bq,Prokushkin:1998vn}. For
$0<  \mu=\lambda < 1$ --- this is the case of relevance since the 't~Hooft coupling 
is by construction between $0< \lambda <1$ ---  the mass therefore
lies in the window where two quantisations are possible \cite{Klebanov:1999tb}. 
The proposal of \cite{Gaberdiel:2010pz} was then that 
one of the scalars is quantised in the standard way (+), whereas the other is quantised in the 
alternate way ($-$). The corresponding primary fields in the dual CFT then have conformal
dimensions equal to $h_{\pm}={1\over 2}(1\pm \lambda)$; these are precisely
the conformal dimensions of the `primitive' representations of the minimal model CFT
in the 't~Hooft limit.

The symmetry algebras of the $\hs{\mu}$ higher spin theory on AdS$_3$, as well as
the 't~Hooft limit of the minimal model CFTs, are both ${\cal W}_{\infty}$ algebras,
but a priori, it is not at all obvious whether they are the {\em same} ${\cal W}_\infty$ 
algebra. This issue was first raised in \cite{Gaberdiel:2011wb}, see also 
\cite{Gaberdiel:2011zw}, and then finally resolved in  \cite{Gaberdiel:2012ku}:
There is a unique way of `quantising' the asymptotic symmetry algebra of the
higher spin theory (that is initially 
a commutative Poisson algebra). The resulting quantum algebra ${\cal W}_{\infty}[\mu]$
exhibits a non-trivial equivalence that implies, among other things, that 
${\cal W}_{\infty}[\lambda]$ agrees indeed with the 't~Hooft limit of the $\W_N$ algebras.
In fact, the equivalence holds also for finite $N$ and $k$ (and hence finite $c$): 
the $\W_{N,k}$ minimal model algebra at central charge $c=c_{N,k}$, see (\ref{coscent}),
is equivalent to the $\w{\lambda}$ algebra at the same value of the central charge 
and with $\lambda$ given by (\ref{thftdef}); this will be reviewed in Sec.~4.

Given the detailed understanding of the $\w{\mu}$ algebra for arbitrary $\mu$ and $c$,
it is then also possible to analyse the semi-classical (large $c$) behaviour of its representations 
at fixed  $\mu$. In particular, one can study the two `primitive' 
coset representations (that correspond to the two quantisations of the massive scalar
field, from above) for fixed $N$ and large $c$. As it turns out, the two representations
behave rather differently in this limit: while the conformal dimension $h_+$ remains finite,
$h_-$ is proportional to $c$. This suggests that the AdS dual of the 
$h_-$ primary should not be thought of as a perturbative massive scalar field with
alternate boundary conditions, but rather as a non-perturbative state  \cite{Gaberdiel:2012ku}.
This point of view also ties in nicely with the fact that the higher spin theory possesses
a large number of semi-classical `conical defect' solutions \cite{Castro:2011iw} that are 
in one-to-one correspondence with the closely related `light' states of the coset CFT.
The picture that emerges from these considerations  \cite{Gaberdiel:2012ku, Perlmutter:2012ds}  
is that the bulk AdS theory should be thought of as a $\hs{\lambda}$ theory coupled to 
{\it one} complex scalar field (dual to $h_{+}$).
Other states, including those dual to  $h_{-}$ and the `light' states, 
are to be viewed as conical defects (and their generalisations) bound with 
perturbative quanta  \cite{Perlmutter:2012ds}; all of this will be discussed in Sec.~5. 

There are various aspects of this proposal that can be checked in some detail. In particular,
one can show that the perturbative spectrum of the higher spin AdS theory matches
exactly with the `perturbative' part of the CFT spectrum, i.e.\ with those states that appear 
in multiple OPEs of the $h_+$ primary (and its conjugate). This calculation represents a 
highly non-trivial consistency check on the proposal, and will be explained, 
in some detail, in Sec.~5.3.
Further checks, including the comparison of correlation functions as well as 
the calculation of the black hole entropy of \cite{Kraus:2011ds} from the dual CFT point of 
view \cite{Gaberdiel:2012yb}
--- for a review about the construction of black hole solutions for these theories see 
\cite{AGKP} --- are discussed in Sec.~6. In Sec.~7 we summarise the generalisations
of the duality conjecture to the orthogonal groups, as well as to the case
with ${\cal N}=2$ supersymmetry. Finally, Sec.~8 outlines some of the possible lines of 
future development of this fruitful subject.

\section{The Ingredients}

In this section we briefly review the basic ingredients that go into the duality, namely, 
higher spin theories on AdS$_3$ on the one hand (see Sec.~2.1), and 
the coset conformal field theories in two dimensions on the other (see Sec.~2.2).

\subsection{The Higher Spin Theory}

Higher spin gauge fields in AdS$_3$ are relatively simple compared to their higher 
dimensional counterparts. (The general Vasiliev approach to constructing higher spin theories in diverse 
dimensions and its relevance for the AdS/CFT correspondence
is reviewed elsewhere in this volume, for instance in the articles by Giombi and Yin \cite{Giombi:2012ms}
and Vasiliev \cite{Vasiliev:2012vf}.)
The basic reason is that these fields, just like gravity, do not contain propagating degrees of freedom in three dimensions. 
Thus their bulk dynamics is topological 
and the only states come from boundary degrees of freedom generalising the Brown-Henneaux 
states of pure AdS$_3$ gravity. The precise higher spin theory that will be dual to the 
$\W_N$ minimal models will, however, have bulk propagating degrees of freedom coming 
from a scalar.  The mass as well as couplings of this scalar are determined by the higher 
spin symmetry algebra. 

Below, we will first review the Chern-Simons construction for pure gravity in AdS$_3$, and then
explain how it can be generalised to higher spin \cite{Blencowe:1988gj}. 
After a discussion of the higher spin 
symmetry algebra we will also mention how the scalar field can be coupled.  

\subsubsection{Review of Pure Gravity}

Recall that the Einstein equations of pure gravity in AdS$_3$ can be written in 
Chern-Simons form \cite{Achucarro:1987vz,Witten:1988hc}. In order to see this,
let us work with the vielbein formalism, where the basic variables are the 
dreibein $e^a_\mu$ and the spin connection $\omega^{bc}_{\mu}$. 
Dualising the spin connection as 
$\omega^a_\mu = - \frac{1}{2} \, \epsilon^a_{bc} \, \omega^{bc}_{\mu}$, the Einstein
equations take the form (in the following we work in form language, and hence
drop the explicit spacetime indices)
\begin{equation}\label{ein1}
R^a \equiv d \omega^a + \frac{1}{2} \epsilon^a_{bc} \omega^b \wedge \omega^c \ =  \ 
\frac{1}{2 \ell^2} \, \epsilon^a_{bc} \, e^b \wedge e^c \ ,
\end{equation}
where $\ell$ is the AdS radius (which will often be set equal to one). In addition, we have 
the condition that the torsion vanishes
\begin{equation}\label{ein2}
T^a \equiv de^a + \epsilon^a_{bc} \omega^b \wedge e^c \ = \  0 \ .
\end{equation}
We now want to obtain these two equations from a Chern-Simons point of view. 
To see how this goes we recall that the isometry group of AdS$_3$ is  
${\rm SO}(2,2) \cong {\rm SL}(2,\mathbb{R}) \times  {\rm SL}(2,\mathbb{R})$. Let us
introduce the fields
\begin{equation}\label{vielconn}
A^a = \omega^a + \frac{1}{\ell}\, e^a \ , \qquad
\bar{A}^a = \omega^a - \frac{1}{\ell}\, e^a \ ,
\end{equation}
that transform in the adjoint representation with respect to the two 
${\rm SL}(2,\mathbb{R})$ factors. Thus both $A^a$ and $\bar{A}^b$ take
values in the Lie algebra $\mathfrak{sl}(2)$, and we can consider the Chern-Simons action 
\begin{equation}\label{CS}
S= S_{\rm CS}[A] - S_{\rm CS}[\bar{A}] \quad \hbox{with} \quad
S_{\rm CS}[A] = \frac{\hat{k}}{4\pi} \int {\rm Tr} \Bigl( A \wedge dA 
+ \frac{2}{3} A \wedge A \wedge A \Bigr) \ .
\end{equation}
It was observed in \cite{Achucarro:1987vz} that the flatness conditions 
$F^a\equiv dA^a + \epsilon^a_{bc} A^b \wedge A^c =  0$ and $\bar{F}^a=0$ that arise
as equations of motion from (\ref{CS}) are in fact equivalent
to the Einstein equations of pure gravity (\ref{ein1}) and (\ref{ein2}). In a similar vein, it was
shown in \cite{Witten:1988hc} that 
the Chern-Simons action (\ref{CS}) reduces, up to some boundary terms, to the Einstein-Hilbert 
action (with negative cosmological constant) provided we identify
\be
\hat{k} = \frac{ \ell}{4 G} \ ,
\ee
where $G$ is Newton's constant. 
We should stress that this identification requires that we choose 
appropriate boundary conditions for the gauge fields.\footnote{Indeed, without
imposing any additional boundary conditions, we would conclude that the field theory living on the
boundary would be a WZW model based on $\mathfrak{sl}(2)$, and this is clearly
{\em not} the conformal field theory dual to pure gravity in AdS$_3$.}  
The precise form of the boundary conditions will be explained in Sec.~3.2.

\subsubsection{Spin 3 and Higher}

Next we want to discuss the generalisation of the above analysis to higher spin theories. 
In three dimensions it is actually possible to define 
consistent higher spin theories containing only a finite number of
spin fields; the simplest example is the theory that contains, in addition to the graviton,
a single field of spin $s=3$. It is simply obtained from the above description by
replacing $\mathfrak{sl}(2)$ by $\mathfrak{sl}(3)$. This is to say, we consider
the Chern-Simons theory of the form (\ref{CS}), where now the gauge fields $A$ and
$\bar{A}$ take values in the Lie algebra $\mathfrak{sl}(3)$ \cite{Henneaux:2010xg,Campoleoni:2010zq}. 
In order to relate this
Chern-Simons theory to a higher spin theory we need to identify the `gravitational'
subalgebra $\mathfrak{sl}(2)\subset \mathfrak{sl}(3)$. The 
most natural choice\footnote{Other choices appeared in the analysis of 
\cite{Ammon:2011nk, Castro:2011fm,Castro:2012bc}.} is to take $\mathfrak{sl}(2)$ to be the principal 
embedding. This essentially means that (the adjoint of) $\mathfrak{sl}(3)$ decomposes as 
\be\label{sl3d}
\mathfrak{sl}(3) = \mathfrak{sl}(2) \oplus {\bf 5} \ ,
\ee
where ${\bf 5}$ denotes the $5$-dimensional $j=2$ representation of $\mathfrak{sl}(2)$.
These components of the two $\mathfrak{sl}(3)$ gauge fields correspond to generalised 
vielbein and connection $1$-forms $e^{ab}$ and
$\omega^{ab}$, respectively, that are symmetric and traceless in the $a,b$ indices
and generalise (\ref{vielconn}). 

In this case, it was shown in \cite{Campoleoni:2010zq} that the resulting equations of
motion of the Chern-Simons theory reduce, at the linearised level,
to the Fronsdal equations \cite{Fronsdal:1978rb}, characterising
a massless spin $s=3$ gauge field on AdS$_3$. Indeed, at the linearised level, 
the generalised vieibeins $e^{ab}_{\mu}$ are related to the symmetric rank three tensor  field
$\phi_{\mu\nu\rho}$ in the Fronsdal formulation as
\be\label{frondict}
\phi_{\mu\nu\rho} \sim {\rm Tr}(e_{(\mu}^{ab}\bar{e}_{\nu a}\bar{e}_{\rho) b}) \ ,
\ee
where $\bar{e}_{\nu a}$ are the background vielbeins for the AdS metric. This demonstrates 
that Chern-Simons theory based on $\mathfrak{sl}(3)$ indeed describes
spin $3$ gravity on AdS$_3$. 

The above construction can be generalised by replacing the gauge group in the Chern-Simons
theory by  $\mathfrak{sl}(N)$ (where  the gravitational $\mathfrak{sl}(2)$ is principally embedded).
The analogue of (\ref{sl3d}) is now
\be\label{slnd}
\mathfrak{sl}(N) = \mathfrak{sl}(2) \oplus {\bf 5} \oplus {\bf 7} \oplus \cdots\oplus {\bf (2N-1)} \ ,
\ee
where the representation of dimension $(2s-1)$ corresponds to the spin $s$ field,  which
is described by generalised vielbein and connection $1$-forms  
$e^{a_1\ldots a_{s-1} }$, $\omega^{a_1\ldots a_{s-1}}$ 
(whose $a_i$ indices are symmetric and traceless), respectively; 
thus the resulting higher spin gauge theory has spin fields of spin $s=2,3, \ldots ,N$. 
At the linearised level, we again have a generalisation of (\ref{frondict}) relating these generalized vielbeins to the Fronsdal fields.  
For more details, we refer the reader to \cite{Campoleoni:2010zq}. 

In all of these cases the higher spin theory is the sum of two Chern-Simons 
terms as in (\ref{CS}) with equal and opposite levels. One can also consider a 
parity violating version of the theory, where the two levels are different
\cite{Chen:2011vp, Bagchi:2011vr}. One needs then to impose the zero torsion condition additionally 
through a Lagrange multiplier term. As a consequence, 
this theory turns out to have propagating modes 
\cite{Bagchi:2011vr, Bagchi:2011td, Chen:2011yx}.

\subsubsection{The Underlying Algebra of the Higher Spin Theory}

The higher spin theories we are primarily interested in are a generalisation of the above 
$\mathfrak{sl}(N)$ theories. They have one massless higher spin field for each 
spin $s=3,4,5,\ldots$. These generalisations 
can be  constructed by considering the Chern-Simons theory \cite{Blencowe:1988gj} 
based on the infinite dimensional
Lie algebra $\hs{\mu}$. Let us first describe the structure of this 
Lie algebra in some detail, following
 \cite{Feigin,Bordemann:1989zi,Bergshoeff:1989ns,Fradkin:1990qk}.

Consider the quotient of the universal enveloping algebra $U(\mathfrak{sl}(2))$ by the ideal 
generated by $(C^{\mathfrak{sl}}-\frac{1}{4}(\mu^2-1) {\bf 1})$, 
\be\label{idealcon}
B[\mu] = {U(\mathfrak{sl}(2))\over \langle C^{\mathfrak{sl}} - \frac{1}{4}(\mu^2-1){\bf 1} \rangle} \ .
\ee
Here $C^{\mathfrak{sl}}$ is the quadratic Casimir of $\mathfrak{sl}(2)$; if we denote the 
generators of $\mathfrak{sl}(2)$ by  $J_0, J_\pm$ with commutation relations
\be
{}[J_+,J_-] = 2 J_0 \ , \qquad [J_\pm,J_0] = \pm J_\pm  \ ,
\ee
then $C^{\mathfrak{sl}}$ is given by 
\be\label{cas}
C^{\mathfrak{sl}} \equiv J_0^2 - \frac{1}{2} (J_+ J_- + J_- J_+)  \ .
\ee
A basis for $B[\mu]$ as a vector space can be described as follows. There is one 
zero letter word, namely the identity generator ${\bf 1}  \equiv V^1_0$ of the universal
enveloping algebra. Then there are three one-letter words, namely 
\be
V^2_1 = J_+ \ , \qquad V^2_0 = J_0 \ , \qquad V^2_{-1} = J_-  \ ,
\ee
five $2$-letter words, since the linear combination described by the Casimir (\ref{cas})
is proportional to ${\bf 1}$ in $B[\mu]$; we may denote them by 
\be\label{V3}
\begin{array}{ll}
V^3_2  = J_+ J_+ \ ,  \quad & \hspace*{-3cm} V^3_1 = J_0 J_+ + \frac{1}{2} J_+\ , \\
V^3_0 = \frac{1}{3} \bigl( J_{-} J_{+} + J_0 + 2 J_0 J_0 \bigr) 
\cong J_0 J_0 - \frac{1}{12}(\mu^2-1) & \\ 
V^3_{-1} =  J_{-} J_0 + \frac{1}{2} J_{-} \ , \quad &  \hspace*{-3cm}
V^3_{-2} = J_{-} J_{-} \ .
\end{array}
\ee
Continuing in this manner one finds that there are 
$2s+1$ different $s-1$ letter words, which we may define to be 
\be\label{env}
V^s_n = (-1)^{s-1-n} \frac{(n+s-1)!}{(2s-2)!} \, 
 \Bigl[ \underbrace{J_- , \dots [J_-, [J_-}_{\hbox{\footnotesize{$s-1-n$ terms}}}, J_+^{s-1}]]\Bigr] \ ,
\ee
where $|n|\leq s-1$. Thus we have a basis for the full vector space $B[\mu]$ 
given by $V^s_n$ with $s=1,2,\ldots$ and $|n|\leq s-1$. 

The vector space $B[\mu]$ actually defines an associative algebra,
where the product  $\star$ is the one inherited from the universal enveloping algebra, i.e.\ 
is defined by concatenation; this is what is sometimes called the `lone-star product' 
in the literature. We can thus turn $B[\mu]$ into a Lie algebra by 
defining the commutator of two generators $X,Y\in B[\mu]$ to be
\be
[X,Y]  = X \star Y - Y \star X \ .
\ee
On $B[\mu]$ we can define an invariant bilinear trace \cite{Vasiliev:1989re} via
\be\label{quadform}
\tr(X \star Y) =  \left. X\star Y \right|_{J_a=0} \ ,
\ee
i.e.\ by retaining only the term proportional to ${\bf 1}=V^1_0$ (after rewriting the product
in terms of the generators $V^s_n$). One easily checks that 
this trace is symmetric. Thus, the commutator of two elements in $B[\mu]$ does not involve
${\bf 1}$, and hence, as a Lie algebra, $B[\mu]$ decomposes as 
\be\label{Bdecomp}
B[\mu] = \mathbb{C} \oplus \hs{\mu} \ ,
\ee
where the vector corresponding to ${\mathbb C}$ in (\ref{Bdecomp}) is the identity 
generator ${\bf 1}$ of the universal enveloping algebra, and a basis of the Lie algebra
\hs{\mu}, thus defined, is given by $V^s_n$ with $s=2,\ldots$ and $|n|\leq s-1$. The generators with $s=2$
define an $\mathfrak{sl}(2)$ subalgebra, with respect to which the generators $V^s_n$ transform 
in the $(2s-1)$-dimensional representation
\be\label{twoaction}
[V^2_m, V^s_n] = \bigl(-n + m(s-1) \bigr) V^s_{m+n} \  .
\ee
We thus conclude that the bulk fields associated to $V^s_n$ have spacetime spin $s$. The 
Chern-Simons theory based on $\hs{\mu}$
therefore describes a higher spin theory with massless spin fields of spin $s=2,3,4,\ldots$. 
\smallskip

Let us analyse the structure of the Lie algebra $\hs{\mu}$ in a little more detail. 
Using (\ref{V3}), the first few commutators are for example
\be\label{threeaction}
\begin{array}{ll}
\hspace*{-1cm} {}[V^3_2,V^3_1] = 2 \, V^4_3 \qquad 
&[V^3_2,V^3_0] = 4\, V^4_2 \\
\hspace*{-1cm} {}[V^3_2,V^3_{-1}] = 6\, V^4_1 - \frac{1}{5} (\mu^2-4) \, V^2_1 \qquad
&[V^3_2,V^3_{-2}] = 8\, V^4_0 - \frac{4}{5} (\mu^2-4)\, V^2_0 \ .
\end{array}
\ee
A closed formula for all commutation relations is known \cite{Pope:1989sr}, 
see e.g.\ eq.~(A.1) in \cite{Gaberdiel:2011wb}. 
Note that the commutators (\ref{threeaction}) suggest that, for $\mu=2$, the Lie algebra
generated by $V^s_n$ with $s\geq 3$ form a proper subalgebra of \hs{\mu}. In fact, 
this is a special case of a more general phenomenon. If $\mu=N$ with integer $N\geq 2$,
then the quadratic form (\ref{quadform}) degenerates \cite{Vasiliev:1989re,Fradkin:1990qk},
\be
\tr(V^s_m V^r_n) = 0  \quad \mbox{for}  \quad s>N \ .
\ee
This implies that an ideal $\chi_N$ appears, consisting of all generators $V^s_n$ 
with $s>N$.  Factoring over this ideal truncates to the finite-dimensional
Lie algebra $\mathfrak{sl}(N)$,
\be\label{slN}
\hs{\mu=N}/\chi_N  \cong \mathfrak{sl}(N)  \quad (N\geq 2) \ .
\ee
Thus we can think of \hs{\mu} as being the continuation of $\mathfrak{sl}(N)$ to non-integer
$N$.  This relation will be important in the following. 

In summary, we therefore have a one-parameter family of higher spin theories on AdS$_3$
that are described by a Chern-Simons theory based on the Lie algebra
$\hs{\mu} \times \hs{\mu}$. The classical theory reduces to a higher spin theory  
with a finite number of spins only when we take the parameter $\mu$ to equal a positive 
integer greater than or equal to $2$; in fact, if $\mu=N$, then the theory becomes
the $\mathfrak{sl}(N)\times \mathfrak{sl}(N)$ higher spin theory described in the previous subsection.

\subsubsection{Coupling to Scalar Fields}

Unlike in higher dimensions, in three dimensions 
the scalar field is not part of the higher spin multiplet and its presence in the theory is 
optional. The theory with a scalar field becomes considerably more complicated 
than the pure higher spin theory since the scalar field carries 
propagating degrees of freedom. 

The full set of interactions of the scalar with the higher spin fields is difficult to write out explicitly
\cite{Prokushkin:1998bq, Prokushkin:1998vn}. However, the interactions at the linearised level 
are relatively simple (see, for instance, \cite{Ammon:2011ua}). The scalar field $C_0(x)$ is the 
part proportional to the identity of a field $C(x)$ which takes values in the Lie algebra $B[\mu]$ 
(see (\ref{Bdecomp})). The latter obeys the linearised field equation
\be\label{Clin}
dC+A\star C-C\star\bar{A}=0 \ ,
\ee
where $A,\bar{A}$ are the $\hs{\mu}$ gauge fields introduced in the previous section. 
When expanded around the AdS vacuum, these field equations imply that the 
scalar obeys the 
Klein-Gordon equation with mass $M^2=-1+\mu^2$. (Here we have set the AdS radius
$\ell=1$.)  Note that for any real value of 
$\mu$ this is  above the Breitenlohner-Freedman bound \cite{Breitenlohner:1982jf} 
$M^2_{\rm BF}=-1$. 
One can also work out the cubic and higher couplings of the scalar field, 
see \cite{Prokushkin:1998bq, Prokushkin:1998vn,Chang:2011mz,Ammon:2011ua,Chang:2011vk},
but we will not go into the details here. 

\subsection{The $\W_N$ Minimal Model CFTs} \label{sec:minmod}

The CFTs we are interested in are the so-called ${\cal W}_N$ minimal models \cite{Fateev:1987zh}. They have 
higher spin conserved currents whose charges form an extended {\it global} symmetry of 
the CFT  --- in contrast to the higher spin {\it gauge} symmetry of the bulk AdS theory 
described in 
the previous subsection. This is, of course, to be expected from the point of view of the 
AdS/CFT correspondence where gauge fields in the bulk AdS couple to conserved currents 
in the boundary theory. 

Interacting 2d conformal field theories with conserved currents $W^{(s)}(z)$ with 
spin $s \geq 3$ were first constructed by Zamolodchikov \cite{Zamolodchikov} and
called $\W$-algebras. They define a new class of chiral algebras beyond the
more familiar cases of (super-)Virasoro/Kac-Moody algebras. In the following
we shall describe one route towards these theories, namely by explaining 
the construction of the $\W_N$ minimal models via the coset construction.
We shall also review their 
spectrum of primary operators, and sketch the structure of the 
associated partition function.

\subsubsection{The Coset Construction}

The $\W_N$ minimal models 
are most easily described in terms of a coset  \cite{Bais:1987zk}
\begin{equation}\label{gencos}
\frac{{{\rm SU}(N)}_k \otimes {\rm SU}(N)_1 }{ {\rm SU}(N)_{k+1}} \ ,
\end{equation}
which is a special instance of the general $G/H$ coset construction \cite{Goddard:1986ee}. 
In our case, this means that we consider a WZW theory based on the group 
$G={\rm SU}(N)\otimes {\rm SU}(N)$ in which we gauge the diagonal subgroup 
$H={\rm SU}(N)$. 
The stress tensor of the coset theory equals
\be\label{cosetstress}
T_{G/H}=T_G-T_H \ ,
\ee
where the individual stress tensors $T_G$ and $T_H$ are given by the usual Sugawara 
construction, i.e.\ in terms of bilinears of the currents. The stress tensors $T_{G/H}$ and $T_H$ 
have non-singular OPE's with each other by construction. We can therefore decompose the 
Hilbert space ${\cal H}_G$ (or more particularly, the affine representation space 
${\cal H}^{(\Lambda)}_G$ corresponding to a highest weight representation $\Lambda$) into 
representations of $H$ as 
\be\label{hilbdecomp}
{\cal H}^{(\Lambda)}_G=\bigoplus_{\Lambda^{\prime}}
\Bigl({\cal H}^{(\Lambda, \Lambda^{\prime})}_{G/H}
\otimes {\cal H}^{(\Lambda^{\prime})}_H\Bigr)\ .
\ee
The multiplicity spaces ${\cal H}^{(\Lambda, \Lambda^{\prime})}_{G/H}$ then define the 
Hilbert space of the coset theory, and the corresponding operators 
commute with the $H$ currents (i.e.\ have a non-singular OPE with them).  

It follows from (\ref{cosetstress}) that the central charge of the coset stress tensor $T_{G/H}$
equals
\be\label{centch}
c_{G/H}=c_G-c_H\ .
\ee
For our particular coset (\ref{gencos}) this leads to 
\begin{eqnarray}\label{cNk} 
c_{N,k} &=& (N^2-1)  \Bigl[ \frac{k}{N+k} + \frac{1}{N+1} - \frac{k+1}{N+k+1}  \Bigr] \ , \cr
& = &  (N-1) \Bigl[1- \frac{N(N+1)}{(N+k)(N+k+1)}\Bigr] \leq  (N-1) \ .
\end{eqnarray}
We will at times  also use the notation $p=N+k \geq (N+1)$. Note that for $N=2$, 
(\ref{gencos}) agrees exactly with the original coset construction of 
\cite{Goddard:1986ee}, that describes the familiar unitary series of the 
Virasoro minimal models with
\be
c_{2,k}= 1- \frac{6}{p(p+1)}\ , \qquad p=k+2 \ .
\ee

For general $N$, the coset theory (\ref{gencos}) with the smallest value of $k=1$, i.e.\ 
$p=N+1$, has central charge
$c=\frac{2(N-1)}{N+2}$, and can alternatively be realised  in terms of 
${\mathbb Z}_N$ parafermions \cite{Fateev:1985mm}. The other extreme case corresponds to 
$p\rightarrow \infty$ (taking $k\rightarrow \infty$ while keeping $N$ finite), where $c = (N-1)$, and
the symmetry algebra is equivalent to the Casimir algebra of the $\mathfrak{su}(N)$ affine algebra 
at level  $k=1$ \cite{Bais:1987dc,Bais:1987zk}. The Casimir algebra consists of all 
$\mathfrak{su}(N)$ singlets in the affine vacuum representation of the affine algebra. 
Since the affine algebra is at  level one, it can be realised in terms of $(N-1)$ free 
bosons; thus in this limit the coset model can be described as a singlet sector of a free
(boson) theory \cite{Gaberdiel:2011aa}. 

\subsubsection{Higher spin Currents}

The cosets (\ref{gencos}) are the simplest examples of interacting CFTs which have 
(for $N \geq 3$) conserved currents of spin $s> 2$. We now describe an explicit
method for constructing these higher spin currents. Actually, this procedure applies to the 
more general cosets of the form
\be\label{gencoset}
\frac{G}{H}=\frac{{\rm SU}(N)_k\otimes {\rm SU}(N)_l}{{\rm SU}(N)_{k+l}} \ .
\ee
Let us consider the cubic combination of currents 
\begin{eqnarray}\label{w3comb}
W^{3}(z) & \propto & 
d_{abc}\biggl(a_1(J^a_{(1)}J^b_{(1)}J^c_{(1)})(z)+a_2(J^a_{(2)}J^b_{(1)}J^c_{(1)})(z) \cr
&& \qquad + a_3(J^a_{(2)}J^b_{(2)}J^c_{(1)})(z)+a_4(J^a_{(2)}J^b_{(2)}J^a_{(2)})(z) \biggr) \ ,
\end{eqnarray}
where $d_{abc}$ is the totally symmetric cubic invariant of 
$\mathfrak{su}(N)$ which is present for $N \geq 3$, while $a_j$ are initially free parameters. 
The currents $J^a_{(1)}, J^a_{(2)}$ refer to the $\mathfrak{su}(N)$ currents in the two factors 
in the numerator of the coset. 
The OPE of any of the four independent terms on the RHS with the diagonal current $(J^a_{(1)}+J^a_{(2)})$
will generate singular terms of the kind $d_{abc}J^b_{(i)}J^c_{(j)}$ with $i,j\in\{1,2\}$. 
Since there 
are only three such terms (since $d_{abc}$ is symmetric), we can choose the $a_j$ 
such that the resulting 
$W^{3}(z)$ has a nonsingular OPE with $(J^a_{(1)}+J^a_{(2)})$. 
Thus it defines a chiral current of weight and spin three in the coset theory. The explicit 
expressions for the coefficients can be found, for instance, in eqs.~(7.42) and (7.43) of 
\cite{Bouwknegt:1992wg}.

Since $\mathfrak{su}(N)$ has independent invariant symmetric tensors for each rank $s$ 
with $s\leq N$ --- these are the independent Casimirs of $\mathfrak{su}(N)$ --- a 
similar construction exists for each spin $s \leq N$. Indeed, the analogue of the ansatz
(\ref{w3comb}) contains now $(s+1)$ independent terms, and the OPE with the diagonal
$(J^a_{(1)}+J^a_{(2)})$ current generates a singular term involving $s$ distinct combinations 
of currents. By choosing the $(s+1)$ coefficients suitably, we can then arrange to have 
one combination which has a non-singular OPE with the diagonal current. Thus we obtain 
one such field $W^{s}(z)$ for every spin $s \leq N$. 

This construction works for general $l$ and $k$ in (\ref{gencoset}). What is special about
taking one of the level, say $l=1$, is that the OPEs of the $W^{s}$  close among 
themselves. (The additional fields that are generated in these OPEs for general $l$
become null for $l=1$ and hence decouple, see e.g.\  \cite{Bouwknegt:1992wg}.)
The resulting algebra of the 
$W^{s}(z)$ defines the  $\W_N$ algebra that is of primary interest to us here.

\subsubsection{Minimal Model Primaries}

The above higher spin currents are in the vacuum sector ($\Lambda=\Lambda^{\prime}=0$ 
in the notation of (\ref{hilbdecomp})) of the coset Hilbert space since they are built purely 
from the currents $J^a_{(1)}$ and $J^b_{(2)}$, and are singlets with respect to the diagonal
zero mode action. The other states of the theory (\ref{gencos}) fall into non-trivial highest 
weight representations of the coset algebra. As is clear from (\ref{hilbdecomp}), a
general representation is parametrised by 
taking $\Lambda=\rho\oplus \mu$, where $\rho$ is a highest weight representation 
(hwr) of $\mathfrak{su}(N)_k$, while $\mu$ is a hwr of $\mathfrak{su}(N)_1$, and 
$\Lambda'=\nu$, where $\nu$ is a hwr of 
$\mathfrak{su}(N)_{k+1}$; thus the most general coset representations are labelled 
by $(\rho,\mu;\nu)$.\footnote{It is important to note though that the states in the coset do 
{\it not} transform under any non-trivial representations of $\mathfrak{su}(N)$.}
Actually, only those combinations are allowed for which $\nu$ appears in the decomposition 
of  $(\rho\oplus \mu)$ under the action of $\mathfrak{su}(N)_{k+1}$. The relevant selection 
rule is simply that 
\begin{equation}
\rho+ \mu - \nu   \in  \Lambda_R \ ,
\end{equation}
where here $\rho$, $\mu$ and $\nu$ are thought of as weights of the finite-dimensional
Lie algebra $\mathfrak{su}(N)$, and $\Lambda_R$ is the corresponding root lattice. 
In addition, there are field  identifications: the two triplets 
\begin{equation}\label{fid}
(\rho,\mu;\nu) \cong (A\rho, A\mu; A\nu) \ ,
\end{equation}
define the same highest weight representation of the coset algebra, provided that $A$ is an 
outer automorphism of the affine algebra $\mathfrak{su}(N)_{k}$. The group  of outer
automorphisms of $\mathfrak{su}(N)_{k}$ 
is ${\mathbb Z}_N$ (independent of $k$), and it is generated by the cyclic rotation of the
affine Dynkin labels $l_j$, i.e.\ the map
\begin{equation}\label{afflab}
[l_0;l_1,\ldots ,l_{N-1}] \mapsto
       [l_1;l_2,\ldots ,l_{N-1},l_0] \ ,
\end{equation}
where the first entry is the affine Dynkin label. In this notation, the allowed highest weight 
representations of  $\mathfrak{su}(N)$ at level $k$ are labelled by 
\begin{equation}
P^+_k(\mathfrak{su}(N)) = 
\Bigl\{ [l_0;l_1,\ldots ,l_{N-1}]  \ : \ 
l_j \in {\mathbb N}_0 \ , \quad \sum_{j=0}^{N-1} l_j = k \Bigr\} \ .
\end{equation}
Note that the field identification (\ref{fid}) does not have any fixed points since 
${\mathbb Z}_N$ acts transitively on the highest weight representations of 
$\mathfrak{su}(N)$ at level $k=1$.  

\subsubsection{The Spectrum of Primaries}

It is easy to see that for {\em any} choice of highest weight representations 
$(\rho;\nu)$, there always exists a unique $\mu\in P^+_1(\mathfrak{su}(N))$, such that 
$\rho+\mu-\nu \in \Lambda_R$. Thus we may label the highest weight 
representations of the coset algebra in terms of unconstrained pairs $(\rho;\nu)$ 
and suppress the $\mu$ label since it is completely determined by the other two. 
The labels are still subject to the field identifications 
\begin{equation}\label{fieldidr}
(\rho;\nu) \cong (A\rho;A\nu) \ .
\end{equation}

Since the coset theory has a stress tensor which is the difference of the two stress tensors of the 
mother and daughter theories, the conformal weight of the corresponding highest 
weight representation has the form
\begin{equation}\label{hco}
h(\rho;\nu) =  \frac{C_2(\rho)}{N+k} + \frac{C_2(\mu)}{N+1} - \frac{C_2(\nu)}{N+k+1} + n \ ,
\end{equation}
where $C_2(\sigma)$ is the eigenvalue of the quadratic Casimir
operator of $\mathfrak{su}(N)$. 
Furthermore, $n$ is a non-negative integer, describing 
the `height' (i.e.\ the conformal weight above the ground state) at which the 
$\mathfrak{su}(N)_{k+1}$ primary $\nu$ appears in the representation $(\rho\oplus \mu)$.
Unfortunately, an explicit formula for $n$ is not available, but it is not difficult to work out $n$
for simple examples. 

Alternatively, one may use the Drinfel'd-Sokolov description of these models, see e.g.\ 
\cite{Bouwknegt:1992wg} for more details. In that language  the highest weight representations 
are labelled by $(\Lambda_+,\Lambda_-) \cong (\rho;\nu)$, and the conformal weights equal
\be\label{hexpsimp}
h(\Lambda_+,\Lambda_-)={1\over 2p(p+1)}\, 
\Bigl( \, \biggl| (p+1)(\Lambda_+ +\hat\rho) -p(\Lambda_- + \hat\rho)\biggr|^2 -\hat\rho^2 \Bigr) \ ,
\ee
where $\hat\rho$ is the Weyl vector of $\mathfrak{su}(N)$.
For $N=2$ (the Virasoro minimal models), (\ref{hexpsimp}) reduces to the familiar
formula 
\begin{equation}
h(r,s) = \frac{(r (p+1) - sp)^2 - 1}{4 p (p+1) } = h(p-r,p+1-s)
\end{equation}
with $1\leq r \leq p-1$, $1\leq s \leq p$. Here we have identified 
$\Lambda_+ = \frac{(r-1)}{2}\vec{\alpha}$ and $\Lambda^- =  \frac{(s-1)}{2}\vec{\alpha}$ (with
$|\vec{\alpha}|^2=2$).
\medskip

In the following, the primary where $\nu=[1,0^{N-2}]={\rm f}$ is the fundamental
representation\footnote{Note that the representation of the affine $\mathfrak{su}(N)$ algebra 
has $N$ entries as in (\ref{afflab}). Here and below we will mostly 
drop the affine Dynkin label, and use a description in terms of the usual
($N-1$) Dynkin labels for representations of  the finite dimensional Lie algebra 
$\mathfrak{su}(N)$.}
with $\rho=[0^{N-1}] = 0$ the trivial representation will play an important role. 
Then either (\ref{hexpsimp}) or (\ref{hco})  gives --- in the latter case $\mu={\rm f}$ with $n=0$ 
\begin{equation}\label{h0f}
h(0;{\rm f}) = \frac{C_2({\rm f})}{N+1} - \frac{C_2({\rm f})}{N+k+1} = 
\frac{(N-1)}{2N} \, \Bigl( 1 - \frac{N+1}{N+k+1} \Bigr) \ ,
\end{equation}
where we have used that 
$C_2({\rm f}) = \frac{1}{2}(\Lambda_{\rm f},\Lambda_{\rm f} +2 \hat\rho) = \frac{N^2-1}{2N}$. On
the other hand, for the coset representation with $\rho={\rm f}$ and $\nu=0$,  $\mu$ is
the anti-fundamental representation, $\mu=\bar{\rm f}$, and we get (again with $n=0$) 
\begin{equation}\label{hf0}
h({\rm f};0) = \frac{C_2({\rm f})}{N+k} +  \frac{C_2({\rm f})}{N+1}  = 
\frac{(N-1)}{2N} \, \Bigl( 1 + \frac{N+1}{N+k} \Bigr) \ .
\end{equation}
An example with $n=1$ arises for the case where $\rho=0$ and $\nu={\rm adj}$, the adjoint
representation. Then $\mu=0$ but $n=1$, and we obtain 
\begin{equation}\label{h0adj}
h(0;{\rm adj}) = 1 - \frac{C_2({\rm adj})}{N+k+1} = 1 - \frac{N}{N+k+1}  \ ,
\end{equation}
where we have used that $C_2({\rm adj}) = N$. Finally, the representation with 
$\rho={\rm adj}$ and $\nu=0$ also has $\mu=0$ and $n=1$, and the conformal weight is 
\begin{equation}\label{hadj0}
h({\rm adj};0) = 1 + \frac{C_2({\rm adj})}{N+k} = 1 + \frac{N}{N+k}  \ .
\end{equation}

\subsubsection{Partition Functions}

To determine the complete partition function of the theory, we need to know the full tower of 
descendants for each of the allowed primaries. These descendant states are generated from the 
ground states by the action of the negative Virasoro and higher spin modes, modulo the
null states that decouple. The most efficient way to calculate the corresponding character 
is by using (\ref{hilbdecomp}) since we know the affine characters 
$\chi^{(\rho\oplus\mu)}_G$ and $\chi^{(\nu)}_H$, and hence can read off the character 
of the coset as a branching function.

For the coset we are considering, the branching functions $b_{(\Lambda_+;\Lambda_-)}$ 
are known explicitly,  see e.g.\ eq.\ (7.51) of \cite{Bouwknegt:1992wg}, and given by
\begin{equation}\label{brfn}
b_{(\Lambda_+;\Lambda_-)}(q) =
 {1 \over \eta(q)^{N-1}} \sum_{w \in \hat{W}} 
 \epsilon(w) q^{{1 \over 2 p(p+1)} ( (p+1)w(\Lambda_+ + \hat\rho) 
 - p (\Lambda_- + \hat\rho) )^2}\ ,
\end{equation}
where $\hat\rho$ is the Weyl vector of $\mathfrak{su}(N)$, $p=k+N$ and the sum is over 
the full affine Weyl group $\hat{W}$.

The full partition function is then obtained by putting together the holomorphic and 
anti-holomorphic branching functions in a modular invariant manner. There are 
many non-equivalent ways of doing so. However, we will be concentrating here on the 
simplest choice --- the so-called `charge conjugation' invariant. 
Its partition function is given by
\be\label{diagpart}
Z_{\rm cc}=\sum_{\Lambda_+, \Lambda_-}|b_{(\Lambda_+;\Lambda_-)}(q)|^2.
\ee

\subsubsection{Fusion Rules and Characters}\label{FuCh}

The fusion rules of the coset theory follow directly from the mother and daughter theory. Indeed,
in terms of the triplets $(\rho,\mu;\nu)$ the fusion rules are simply
\begin{equation}
{\cal N}_{(\rho_1,\mu_1;\nu_1)\, (\rho_2,\mu_2;\nu_2)}{}^{(\rho_3,\mu_3;\nu_3)}
  = {\cal N}_{\rho_1 \rho_2}^{(k)\ \rho_3} \,\, {\cal N}_{\mu_1 \mu_2}^{(1)\ \mu_3} \, \,
  {\cal N}_{\nu_1 \nu_2}^{(k+1)\ \nu_3} \ ,
\end{equation}
where the fusion rules on the right-hand side are those of ${\mathfrak g}_k$, 
${\mathfrak g}_1$ and ${\mathfrak g}_{k+1}$, respectively. Note that the fusion rules are 
invariant under the field identification (\ref{fid}). Since the fusion rules of the level one factor
are just a permutation matrix, we can also directly give the fusion rules for the representatives
$(\rho;\nu)$ as
\begin{equation}
{\cal N}_{(\rho_1;\nu_1)\, (\rho_2;\nu_2)}{}^{(\rho_3;\nu_3)}
  = {\cal N}_{\rho_1 \rho_2}^{(k)\ \rho_3} \,\,  {\cal N}_{\nu_1 \nu_2}^{(k+1)\ \nu_3} \ .
\end{equation}
Note that the fusion rules on the RHS for the affine theories are strongly constrained by the 
$\mathfrak{su}(N)$ symmetry -- they are essentially Clebsch-Gordan coefficients. This will 
play an important role when we consider the large $N$ limit of correlation functions. 

This completes our brief review of the $\W_N$ minimal models; further details 
about coset theories in general can be found in \cite{DiFrancesco:1997nk}, and 
various aspects of $\W$-algebras are explained in the review \cite{Bouwknegt:1992wg}.

\section{$\W_\infty$-Symmetries in the Boundary and the Bulk}

In this section we explain the emergence of $\W_\infty$ symmetries
in our context. First, in Sec.~3.1 (see in particular Sec.~3.1.1) we 
describe the $\W_\infty$ algebras that appear in the 2d CFTs of 
free bosons and free fermions. Then, in Sec.~3.2, we show
that closely related $\W_\infty$ algebras appear as the asymptotic
symmetry algebra of higher spin theories in AdS$_3$. The precise
relationship between the two constructions will be the topic of 
Sec.~4. 

\subsection{$\W$-algebras in 2d CFTs}

Unlike in higher dimensions, it is possible to have non-trivial interacting 
quantum field theories in two dimensions which possess conserved currents of spin 
$s > 2$. The Coleman-Mandula argument \cite{Coleman:1967ad}
does not rule out their existence and indeed 
there is a rich collection of  2d (massive)  integrable quantum field theories which have 
higher spin conserved charges. 

The Coleman-Mandula theorem itself applies to theories with an S-matrix and there is an 
assumption about the spectrum having a mass gap \cite{Coleman:1967ad}. Therefore it does not strictly apply to 
conformal field theories. An analogue of the Coleman-Mandula theorem (with some mild 
assumptions) was recently proven for conformal field theories in 3d  
\cite{Maldacena:2011jn}.\footnote{The proof may be generalisable to higher dimensions, see 
\cite{Stanev:2012nq,Zhiboedov:2012bm} for first steps in this direction.}
This theorem shows that (in a theory with a finite number of fields) the correlation functions of 
higher spin currents are necessarily given  by either those in a theory of free fermions, 
or in one of free bosons. Thus, at least the sector of higher spin currents behaves like a free 
theory. 

In two dimensions this theorem does not hold. 
Indeed, as we have seen in Sec.~2.2, the coset conformal theories (that are typically
interacting) have conserved higher spin currents. The resulting $\W$-algebras
are generically non-linear (in contradistinction to the more familiar 
Kac-Moody or Virasoro algebras). This is to say 
quadratic terms in the current modes appear on the RHS of current commutators. The 
OPE of the currents is nevertheless associative and hence Jacobi identities are obeyed. This 
non-linear structure is directly responsible for the fact that the algebra undergoes a non-trivial
deformation at the quantum level (as we shall explain in some detail below, see Sec.~4).
It is also the reason why these theories are much harder
to analyse. In fact, the complete commutation relations have only been written down
explicitly for a few $\W$-algebras involving fields of small spin.

While a large number of different $\W$-algebras have been studied (and there are 
probably many more yet to be discovered) we will  restrict our attention in this review 
mostly to a special class of $\W$-algebras. We will consider the so-called $\W_N$ algebras 
which contain exactly one conserved current $W^{s}(z)$ of integer spin $s=2,3,\ldots N$, 
with the spin two current being the stress tensor $W^{2}(z)\equiv T(z)$. For fixed $N$, 
these algebras are parametrised by the central charge $c$, and for $c=c_{N,k}$, see (\ref{cNk}),
the algebras
coincide with those arising in the coset construction of Sec.~2.2. For $c\geq N-1$, 
the algebras appear also in a family of (generically 
non-unitary) CFTs known as the $A_{N-1}$ Toda theories of which the  Liouville theory is the 
simplest member (for $N=2$). 

The $\W_N$ algebras are in turn a special case of an even more general family of 
$\W$-algebras which will 
play a central role in our analysis and which we will denote by 
$\W_{\infty}[\mu]$. These algebras are parametrised by two labels: the central charge $c$,
as well as the parameter $\mu$. Generically, the algebras are generated by the 
currents $W^{s}(z)$, where $s=2,3,\ldots$ without any bound on $s$. For special
values of $\mu$, however, e.g.\ for $\mu=N\geq 2$, they reduce to the $\W_N$ algebras 
discussed above. There are also simplifications for $\mu=0$ and $\mu=1$, where 
the algebras are closely related to those of free fermions and free bosons, respectively. 

The higher spin algebra corresponding to free bosons and fermions is an analogue of a similar 
algebra in the higher dimensional theories (though extended in two dimensions to chiral and 
anti-chiral currents); the $\W_{\infty}[\mu]$ algebras for $\mu\neq 0,1$, on the other hand,
do not seem to have an analogue in higher dimensions. 
In the following we shall sketch the construction of the $\W_\infty$-algebras for 
the case of free bosons and free fermions; we shall come back to the $\W_\infty[\mu]$
algebra for general $\mu$ in Sec.~4.

\subsubsection{Free Bosons and Free Fermions}

As in higher dimensions, we can write down conserved currents of spin higher than two 
for a system of free bosons or free fermions. The new feature in two dimensions is the 
enlargement to separate holomorphic and anti-holomorphic currents as in the case of the 
stress tensor. 

Consider, for instance, a complex free boson. We can write down conserved currents with 
$s=1,2, \ldots \infty$ \cite{Bakas:1990ry, Pope:1991ig}
\be\label{boson}
W_B^{s}(z) \propto \sum_{k=0}^{s-2}(-1)^k {s-1 \choose k}
{s-1 \choose k+1}\, \partial^{s-k-1}\bar{\phi}\, \, \partial^{k+1}\phi \ .
\ee
For $s=1,2$ this reduces to the familiar charge and energy momentum currents, respectively. The 
combinatorial coefficients are chosen so that $W_B^{s}$ transforms as a quasi-primary under 
the global conformal transformations. The conservation follows from the equation of motion for 
the free theory $\bar{\partial}\partial\phi=0$. It is straightforward to generalise this construction to 
an $N$ component boson $\phi_i$ --- the above currents are  singlet bilinears under the resulting 
global ${\rm SU}(N)$. 

Using the OPE of free bosons it is not difficult to work out the OPE of the currents $W_B^{s}(z)$. 
Schematically one finds
\be\label{wfreeope}
W^{s}\cdot W^{s^{\prime}} \sim W^{s+s^{\prime}-2}+W^{s+s^{\prime}-4}+\cdots 
+ c_s\delta_{s,s^{\prime}} \ ,
\ee
with a central term $c_s \propto c=N$ for the general case of $N$ free bosons. 
Note that the algebra has no non-linear terms. Explicit expressions for the commutation 
relations of the modes of these currents, can, for example, be found in Sec.~3 of 
\cite{Pope:1991ig}. The resulting Lie algebra is related by a nonlinear change of basis to the 
general $\W_\infty[\mu]$ algebra at the special value $\mu=1$ (and $c=N$) 
\cite{Gaberdiel:2011wb}. 
 
Similarly for a free Dirac fermion we can define
\cite{Bergshoeff:1989ns, Bergshoeff:1990yd, Depireux:1990df, Pope:1991ig}
\be\label{ferm}
W_F^{s}(z) \propto \sum_{k=0}^{s-1}(-1)^k {s-1 \choose k}^2\, 
\partial^{s-k-1}\bar{\psi}\, \, \partial^{k}\psi \ ,
\ee
with the $s=1,2$ expressions being the more familiar conserved currents. Again these 
combinations are quasi-primary, and  the OPE also has the schematic form 
as in (\ref{wfreeope}) though the (suppressed) coefficients in front of the individual terms 
are different, see \cite{Pope:1991ig} for explicit expressions. This algebra is believed to be 
related to the $\W_\infty[\mu]$ algebra at $\mu=0$ after an appropriate truncation to the sector 
without the spin one current \cite{Gaberdiel:2011wb}.

\subsection{Asymptotic Symmetries of Higher Spin Theories}

Next we want to explain how similar $\W_\infty$ algebras also appear as asymptotic
symmetry algebras of higher spin gauge theories on AdS$_3$. Recall from 
Sec.~2.1 that these higher spin gauge theories 
can be described in terms of a Chern-Simons theory. In this section, we pay close 
attention to the boundary conditions that need to be imposed in this description, first 
in the case of pure gravity following closely \cite{Banados:1998gg} as reviewed in
\cite{Campoleoni:2010zq}. Then we explain how to generalise this analysis to the case of 
spin $3$ gravity, and how to obtain the asymptotic symmetry algebra \`a la
Brown \& Henneaux \cite{Brown:1986nw}. (Again this follows closely 
\cite{Campoleoni:2010zq}, see also \cite{Henneaux:2010xg}). Finally
we explain the result for the general $\hs{\mu}$ case that was obtained first 
in \cite{Gaberdiel:2011wb} (see also \cite{Campoleoni:2011hg}). 

\subsubsection{Asymptotic Symmetry Analysis for Gravity} \label{sl2}

In order to describe the boundary conditions in the 
${\rm SL}(2,\mathbb{R}) \times  {\rm SL}(2,\mathbb{R})$ Chern-Simons formulation 
of gravity, let us introduce a  basis for $\mathfrak{sl}(2)$
consisting of $L_{0}, L_{\pm 1}$ with  $[L_m,L_n]=(m-n) L_{m+n}$. 
Furthermore, we parametrise the solid cylinder on which the
Chern-Simons theory is defined by $(t,\rho,\phi)$, where $(\rho,\phi)$ are 2d polar
coordinates on the disc, while $t$ is the variable along the length of the cylinder. 
Introducing light-cone variables as 
\be
x^\pm = \frac{t}{\ell} \pm \phi \ ,
\ee
the $1$-form $A^a$ from (\ref{vielconn}) takes the form
\be
A^a = A_\rho^a d\rho + A^a_+ dx^+ + A^a_{-} dx^{-} \ .
\ee
The solid torus has a boundary, and hence the variation of the Chern-Simons 
action includes the boundary term 
\be
\left. \delta S  \right|_{\rm bdy} = - \frac{\hat{k}}{4\pi} \int_{\mathbb{R}\times S^1} dx^+ dx^-
{\rm Tr} \Bigl( A_+ \delta A_{-} - A_{-} \delta A_+ \Bigr) \ . 
\ee
A natural boundary condition that guarantees that this boundary term vanishes is then 
for example
\be
A_- = 0 \qquad \hbox{at the boundary.} 
\ee
Note that this is necessary in order to really reproduce the equations of motion of 
Einstein gravity from the Chern-Simons point of view.

Next we want to characterise the physically inequivalent solutions of the Chern-Simons
theory that are asymptotically AdS$_3$. We can partially fix the gauge by setting
\be\label{solc1}
A_\rho = b^{-1}(\rho)\,  \partial_\rho b(\rho) \ ,
\ee
where $b(\rho)$ is an arbitrary function with values in ${\rm SL}(2,\mathbb{R})$. 
Solving the equations of motion ($F=0$), then leads to 
\be\label{solc2}
A_+ = b^{-1}(\rho)\, a(x^+) \, b(\rho) \ , \qquad A_- = 0  \ .
\ee
The analysis can be done similarly for $\bar{A}$, leading to 
\be\label{sola}
\bar{A}_\rho = b(\rho)\,  \partial_\rho b^{-1}(\rho) \ , \qquad
\bar{A}_+ = 0 \ , \qquad
\bar{A}_{-}=  b(\rho)\, \bar{a}(x^-)\, b(\rho)^{-1} \ ,
\ee
where $b(\rho)$ is the same function as above --- this is necessary 
for the solution to be asymptotically AdS$_3$. In fact, AdS$_3$ is described in this framework
by the solution
\begin{eqnarray}
A_{\rm AdS} & = & b^{-1} \Bigl(L_1 + \frac{1}{4} L_{-1} \Bigr) \, b\, \,dx^+ +  
b^{-1} \partial_\rho b \, d\rho \label{AAdS} \\
\bar{A}_{\rm AdS} & = & - b\,  \Bigl(\frac{1}{4} L_{1} + L_{-1} \Bigr) \, b^{-1} dx^- + 
b \, \partial_\rho b^{-1}  \, d\rho  \ , 
\end{eqnarray}
where 
\be
b(\rho) = e^{\rho L_0} \ .
\ee
Actually, the condition that (\ref{sola}) takes the above form is not quite sufficient to 
obtain an asymptotically AdS solution (in the sense of Brown \& Henneaux \cite{Brown:1986nw}), 
as was discussed in \cite{Banados:1998gg,Campoleoni:2010zq}. In addition we have to require 
that 
\be\label{asym}
\left. (A - A_{\rm AdS}) \right|_{\rm bdy} = 
\left. (\bar{A} - \bar{A}_{\rm AdS}) \right|_{\rm bdy} = 
{\cal O}(1) \ . 
\ee
In particular, this implies that the functions $a(x^+)$ and $\bar{a}(x^-)$ that appear in 
(\ref{solc2}) and (\ref{sola}) are of the form 
\be\label{AAdS2}
a(\phi) = L_1 + l^0(\phi) L_0 + l^{-1}(\phi) L_{-1} \ , \quad
\bar{a}(\phi) = L_{-1} + \bar{l}^0(\phi) L_0 + \bar{l}^1(\phi) L_{1} \ , 
\ee
where $l^0(\phi)$ and $l^{-1}(\phi)$ (as well as their barred cousins) 
are arbitrary functions of $\phi$, and we have set (for simplicity) $t=0$.

Among the asymptotically AdS solutions we should now identify those as
physically equivalent 
that can be related by a gauge transformation that vanishes at the boundary.\footnote{Indeed, 
since 3d gravity is topological, {\em any} two solutions are gauge equivalent, 
and physical degrees of freedom only arise if we are careful about boundary conditions.}
Using only gauge transformations of this type we can set 
$l^0(\phi)=\bar{l}^0(\phi)=0$,  but we cannot change $l^{-1}(\phi)$ and $\bar{l}^1(\phi)$.
Thus we conclude that the space of physically inequivalent asymptotically AdS solutions are 
parametrised by the functions $l^{-1}(\phi)$ and $\bar{l}^1(\phi)$. This space then
carries naturally an action of ${\rm Diff}(S^1)\times {\rm Diff}(S^1)$, corresponding
to the two commuting Virasoro actions  predicted by the analysis 
of Brown \& Henneaux \cite{Brown:1986nw}. (The asymptotic symmetry analysis can also
be carried out in this framework, see \cite{Banados:1998gg}  --- we shall sketch this for the 
case of spin $3$ gravity in the following section.)

\subsubsection{Asymptotic Symmetry Analysis for Spin $3$ Gravity} \label{sl3}

Now we generalise the analysis to the pure higher spin theory containing 
in addition to the spin two graviton a massless spin
three field. As reviewed in Sec. 2.1.2, there is a Chern-Simons description in terms of 
$\mathfrak{sl}(3) \times \mathfrak{sl}(3)$ gauge fields. In this framework, we 
need to discuss how asymptotically AdS solutions can be characterised. 
To be concrete, let us denote the basis elements of the $5$-dimensional subspace 
in (\ref{sl3d}) by $V^3_n$ with $n=-2,-1,0,1,2$; their commutators are then given by
\begin{eqnarray}
{}[L_m, V^3_n]  & = & (2m-n) V^3_{m+n}  \\
{}[V^3_m,V^3_n] & = & (m-n) (2m^2 + 2n^2 - mn - 8) L_{m+n} \ .
\end{eqnarray}
The most general ansatz for the function $a(\phi)$ in eq.~(\ref{solc2}) is then
(analogous statements hold also for $\bar{a}(\phi)$)
\be\label{ans3}
a(\phi) = \sum_{m=-1}^{1} l^m(\phi) L_m +  \sum_{n=-2}^{2} w^m(\phi) V^3_m \ .
\ee
We can characterise the asymptotic boundary condition as in (\ref{asym}), where
$A_{\rm AdS}$ is the solution for which $w^m(\phi)\equiv 0$, and $l^m(\phi)$
is given as in (\ref{AAdS2}), 
\be\label{asym3}
l^1(\phi) = 1 \ , \qquad w^{2}(\phi) = w^1(\phi)=0 \ . 
\ee
By means of gauge transformations that vanish at the boundary we can also set 
\be\label{aux3}
l^0(\phi) = w^0(\phi) = w^{-1}(\phi) = 0 
\ee
and hence the space of physically inequivalent asymptotically AdS solutions is
parametrised by the functions $l^{-1}(\phi)$ and $w^{-2}(\phi)$ (as well as their
right-moving analogues). 
\smallskip

The next step is now to determine the (classical) asymptotic symmetry algebra of this
higher spin theory. Concentrating on the left-moving fields --- the analysis for the right-movers
is analogous --- the most general gauge transformation that preserves the 
gauge (\ref{solc1}) and (\ref{solc2}) is 
\be\label{gaugetrans}
\Gamma(x^+) = e^{-\rho L_0} \, \gamma(x^+) \, e^{\rho L_0} \ ,
\ee
where $\gamma(x^+)$ is an arbitrary Lie algebra valued function; here we have used that
its action on the gauge field is of the form
\be
\delta a = \gamma' + [a, \gamma] \ .
\ee
Let us parametrise $\gamma(\phi)$ as 
\be\label{gauge}
\gamma(\phi) = \sum_{s=2}^{3} \sum_{|n| < s} \gamma_{s,n}(\phi) V_n^s \ ,
\ee
where $V^2_n\equiv L_n$ with $n=0,\pm 1$. Demanding that, after the gauge transformation
(\ref{gauge}), the gauge connection $A$ is still of the form (\ref{ans3}) with (\ref{asym3}) and
(\ref{aux3}), then leads to the recursion equations (from the conditions that the coefficients of 
$L_1\equiv V^2_1$ and $L_0\equiv V^2_0$ are unchanged)
\begin{eqnarray}
\gamma_{2,0} & = & - \gamma_{2,1}'  \label{eq1} \\
\gamma_{2,-1} & = & \frac{1}{2} \gamma_{2,1}'' + \frac{2\pi}{\hat{k}}\, \gamma_{2,1}\, {\cal L} 
+ \frac{4\pi}{\hat{k}}\, \gamma_{3,2} \, {\cal W} 
\end{eqnarray}
where
\be
{\cal L} (\phi) = \frac{\hat{k}}{2\pi}\, l^{-1}(\phi)  \ , \qquad
{\cal W}(\phi) = \frac{6 \hat{k}}{\pi}\, w^{-2}(\phi) \ .
\ee
Similarly, from the requirement that the coefficients of $V^3_{2}$, $V^3_1$, $V^3_0$ and 
$V^3_{-1}$ continue to vanish, we get
\begin{eqnarray}
\gamma_{3,1} & = & - \gamma_{3,2}' \\
\gamma_{3,0} & = & \frac{1}{2} \gamma_{3,2}'' + \frac{4\pi}{\hat{k}}\, \gamma_{3,2} \, {\cal L} \\
\gamma_{3,-1} & = & - \frac{1}{6} \gamma_{3,2}''' - \frac{10 \pi}{3\hat{k}}\, \gamma_{3,2}' {\cal L}
- \frac{4 \pi}{3\hat{k}}\, \gamma_{3,2}\, {\cal L}' \\
\gamma_{3,-2} & = & \frac{1}{24} \gamma_{3,2}'''' + \frac{4\pi}{3\hat{k}}\, \gamma_{3,2}''\, {\cal L}
+ \frac{7\pi}{6\hat{k}} \gamma_{3,2}'\, {\cal L}' + \frac{\pi}{3 \hat{k}} \gamma_{3,2}\, {\cal L}'' \nonumber \\
& & \quad + \frac{4\pi^2}{\hat{k}^2}\, \gamma_{3,2} \, {\cal L}^2 
+ \frac{\pi}{6\hat{k}}\, \gamma_{2,1}\, {\cal W} \ . \label{eql}
\end{eqnarray}
Writing $\epsilon \equiv \gamma_{2,1}$ and $\chi\equiv \gamma_{3,2}$ we then get
altogether the variations
\begin{eqnarray}
\delta_\epsilon \, {\cal L} & = &  \epsilon\, {\cal L}' + 2 \epsilon'\, {\cal L} + \frac{\hat{k}}{4\pi} \epsilon'''
\label{Lconf} \\
\delta_\epsilon \, {\cal W} & = &  \epsilon\, {\cal W}' + 3 \epsilon'\, {\cal W}  \label{Wconf}
\end{eqnarray}
as well as 
\begin{eqnarray}
\delta_\chi \, {\cal L} & = & 2 \chi \, {\cal W}' + 3 \chi' {\cal W} \\
\delta_\chi\, {\cal W} & = & 2 \chi \, {\cal L}''' + 9 \chi' {\cal L}'' + 15 \chi'' {\cal L}' + 10 \chi''' {\cal L} 
+ \frac{\hat{k}}{4\pi} \chi^{(5)} \nonumber \\
& & \quad + \frac{64 \pi}{\hat{k}}\, \bigl(\chi {\cal L} {\cal L}' + \chi' {\cal L}^2  \bigr) \ . \label{bilin}
\end{eqnarray}
Interpreting these variations in terms of charges, we can read off the Poisson brackets of the
associated currents, see \cite{Campoleoni:2010zq} for details. In particular, it follows from 
eq.~(\ref{Lconf}) that ${\cal L}$ plays the role of the stress energy tensor, i.e.\ that the
associated modes satisfy the Virasoro algebra 
\be\label{LL}
i \{ {\cal L}_m,{\cal L}_n \} = (m-n) \, {\cal L}_{m+n} + \frac{c}{12} \, m (m^2-1) \, \delta_{m,-n} \ , \qquad
c = 6 \hat{k}  \ .
\ee
Furthermore, eq.~(\ref{Wconf}) means that ${\cal W}$ is a primary field of conformal weight 
$h=3$, since we have the Poisson brackets
\be\label{LW}
i \{ {\cal L}_M, {\cal W}_n \} = (2m -n ) {\cal W}_{m+n}  \ .
\ee
Finally, the Poisson bracket of the ${\cal W}$ modes with themselves contain bilinear terms
that originate from eq.~(\ref{bilin})
\begin{eqnarray}
\hspace*{-1.5cm} 
i \{ {\cal W}_m,{\cal W}_n \} &  = & - \Bigl[ (m-n ) (2 m^2 + 2n^2 - mn -8) {\cal L}_{m+n} + 
\frac{96}{c} (m-n) \, \Lambda^{(4)}_{m+n} \nonumber \\
& & \qquad + \frac{c}{12} m (m^2-1) (m^2-4)\, \delta_{m,-n} \Bigr] \ , \label{WW}
\end{eqnarray}
where $\Lambda^{(4)}_m \equiv \sum_{n\in\mathbb{Z}} {\cal L}_{n} {\cal L}_{m-n}$. The 
Poisson algebra defined by (\ref{LL}), (\ref{LW}) and (\ref{WW}) is the classical
${\cal W}_3^{\rm cl}$ algebra, which is a well-defined Poisson algebra (in particular
satisfying the Jacobi identity) for any value of $c$. Because of the non-linear term,
the `quantisation' of this algebra, where we replace Poisson brackets by
commutators, is not straightforward since we will have to worry about normal
ordering terms. We will come back to this issue in Sec.~\ref{matching}.

\subsubsection{Asymptotic Symmetry Algebra of $\hs{\mu}$ Chern-Simons Theory}

Next we want to study the asymptotic symmetry algebra of the Chern-Simons theory 
based on $\hs{\mu}\times \hs{\mu}$; this can be done 
in close analogy to the case of $\mathfrak{sl}(3)$ in Sec.~\ref{sl3}. 
The asymptotic boundary condition (\ref{asym}) together with the gauge transformations
that vanish at the boundary now allow one to set the coefficients of all 
Lie algebra generators $V^s_n$ to zero, except for $V^2_1$ (whose coefficient equals $1$), 
as well as $V^s_{-s+1}$ (whose coefficients $\gamma_{s,-s+1}$ 
are the analogues of the functions $l^{-1}(\phi)$ and $w^{-2}(\phi$ from above). The 
requirement that the gauge transformation  (\ref{gaugetrans}) leaves this form of the solution
invariant leads then again to recursion relations analogous to (\ref{eq1}) -- (\ref{eql}). This 
allows one to determine the variations,
i.e.\ the analogues of (\ref{Lconf}) -- (\ref{bilin}), and from them the Poisson brackets of
the fields ${\cal W}^{(s)} \sim \gamma_{s,-s+1}$. For the first few cases this was explicitly
worked out in \cite{Gaberdiel:2011wb}, and it was observed that the answer agrees precisely
with the classical ${\cal W}^{\rm cl}_\infty[\mu]$ algebra that had been obtained before in 
\cite{FigueroaO'Farrill:1992cv,Khesin:1993ww,Khesin:1993ru}.\footnote{Actually, 
one can argue on general grounds \cite{Campoleoni:2010zq} 
that, at least formally,  the asymptotic symmetry algebra is  the Drinfel'd-Sokolov 
reduction  \cite{Drinfeld:1984qv} (see \cite{Dickey} for a review) of the affine algebra
based on \hs{\mu}. Then the identification of the asymptotic symmetry algebra
with ${\cal W}^{\rm cl}_\infty[\mu]$ can also be deduced from the work of \cite{Khesin:1994ey}.}

Explicit closed form expressions for the Poisson brackets
(albeit in a non-primary basis) are known for ${\cal W}^{\rm cl}_\infty[\mu]$,
see \cite{FigueroaO'Farrill:1992cv} or the appendix of \cite{Gaberdiel:2011wb}. Recursion 
relations for the algebra in a primary basis were later given in \cite{Campoleoni:2011hg}.
The algebra ${\cal W}^{\rm cl}_\infty[\mu]$ is generated by the elements ${\cal W}^{(s)}_n$, where 
$s=2,3,\ldots$ and $n\in\mathbb{Z}$. Because of the non-linear terms 
(i.e.\ the analogue of the $\Lambda^{(4)}$ term in (\ref{WW})), it is not 
immediately clear how to turn the Poisson brackets into commutators --- we shall come
back to this point in Sec.~\ref{matching}. However, these difficulties
go away for $c\rightarrow \infty$ (since the non-linear terms are suppressed by
inverse powers of $c$) \cite{Bowcock:1991zk}. In this limit, the generators
${\cal W}^{(s)}_n$ with $|n|\leq s-1$ --- we shall sometimes refer to the corresponding
algebra as the  `wedge algebra' ---  reduce to those of \hs{\mu}. Thus we can think of 
${\cal W}^{\rm cl}_\infty[\mu]$ as an extension of the wedge algebra
\hs{\mu},  `beyond the wedge'. However, at finite $c$ 
(and with the exception of $\mu=1$), even the commutation relations of the wedge
generators acquire non-linear correction terms and thus do not agree with those of \hs{\mu}.
Thus we expect that \hs{\mu} will not be a subalgebra of the quantum
\w{\mu} algebra. As we have mentioned before, $\mu=1$ corresponds to a free
boson theory, and thus the fact that \hs{\mu} with $\mu\neq 1$ is not a genuine 
symmetry of the theory at finite $c$ is the two-dimensional incarnation of the result of 
\cite{Maldacena:2011jn,Maldacena:2012sf}.

\section{Matching the Symmetries} \label{matching}

Next we want to understand the precise relation between the asymptotic 
symmetry algebra ${\cal W}^{\rm cl}_\infty[\mu]$ of the higher spin theory on AdS 
that we have just derived, and the limit algebra of the ${\cal W}_{N,k}$
minimal models. In order to do so it is important to understand how
we can turn the classical Poisson algebra ${\cal W}^{\rm cl}_\infty[\mu]$  into a consistent
quantum algebra $\w{\mu}$; the following analysis follows closely \cite{Gaberdiel:2012ku}.

\subsection{The Quantum Algebra \w{\mu}} 

As we have mentioned before, the main difficulty in replacing the Poisson brackets
by commutators comes from the non-linear terms in the commutation relations. For example, 
naively `quantising' the Poisson brackets of ${\cal W}^{\rm cl}_\infty[\mu]$ leads to 
the commutators
\begin{eqnarray} 
[W^3_m, W^3_n] &= & 2(m-n)W^4_{m+n}+  \frac{N_3}{12}(m-n)(2m^2+2n^2-mn-8)L_{m+n}  
\nonumber \\
& & 
+\frac{8 N_3}{c}(m-n)\, \Lambda^{(4)}_{m+n} 
+ \frac{N_3 c}{144}m(m^2-1)(m^2-4)\delta_{m,-n}  \\
{}[W^3_m, W^4_n] &= & (3m-2n)W^5_{m+n} 
+ \frac{208 N_4}{25 N_3 c}(3m-2n)\, \Lambda^{(5)}_{m+n}
+ \frac{84 N_4}{25 N_3 c}\, \Theta^{(6)}_{m+n} 
\nonumber \\
& & 
- \frac{N_4}{15 N_3}(n^3 - 5m^3 - 3 m n^2 + 5 m^2 n - 9n + 17 m)W^3_{m+n} \ ,
\end{eqnarray}
where $\Lambda^{(4)}\sim L L$, $\Lambda^{(5)}\sim W^3 L$, and 
$\Theta^{(6)}\sim \frac{2}{3} L (W^{3})' - L' W^{3}$, and we have denoted the Virasoro generators
by $W^2_n\equiv L_n$. Furthermore, the structure constants take the form
\begin{eqnarray}
N_3 & = & \frac{16}{5}\, q^2 \, (\mu^2-4)  \label{N3c} \\
N_4 & = & \frac{384}{35}\, q^4 \, (\mu^2-4)\,  (\mu^2-9) \ , \label{N4c}
\end{eqnarray} 
where $q$ is a normalisation parameter, i.e.\ different values of $q$ describe
the same algebra. 
As written, these commutation relations do not satisfy the Jacobi identities 
\be\label{jacobi}
\hspace*{-1.5cm}
{} [L_m,[L_n,W^3_l]] + \hbox{cycl.} 
= [L_m,[W^3_n,W^3_l]] + \hbox{cycl.} = [W^4_m,[W^3_n,L_l]] + \hbox{cycl.} =0 \ ,
\ee
except to leading order in $1/c$. However, we can satisfy the Jacobi identities exactly, 
i.e.\ for arbitrary finite $c$, by (i) defining carefully what we mean by $\Lambda^{(4)}$, 
$\Lambda^{(5)}$, $\Theta^{(6)}$, i.e.\ by specifying the correct `normal ordering prescription';
and (ii) by {\rm modifying} the above commutation relations by $1/c$ corrections. Explicitly,
the correct normal ordered expressions are 
\begin{eqnarray}
\Lambda^{(4)}_n & = & \sum_{p} : L_{n-p} L_p :  + \frac{1}{5} x_n L_n 
 \\
\Lambda^{(5)}_n & = & \sum_{p} : L_{n-p} W^3_p : + \frac{1}{14} \, y_n W^3_n \label{L5} \\
\Theta^{(6)}_n & = & \sum_{p} (\frac{5}{3} p - n) : L_{n-p} W^3_p :  
+ \frac{1}{6}\, z_n W^3_n \ , \label{L6}
\end{eqnarray}
where
\begin{eqnarray}
&x_{2l} = (l+1)(1-l) \ , \qquad  & x_{2l-1} = (l+1) (2-l) \ , \\
& y_{2l} = (l+2)(3-5l) \ , \qquad & y_{2l-1} = 5 (l+1)(2-l) \  , \label{ydef} \\
& z_{2l} = l (l+2)  \ , \qquad & z_{2l-1} = 0\ ,
\end{eqnarray}
and the modified form of the above commutation relations read as
\begin{eqnarray} 
[W^3_m, W^3_n] &= & 2(m-n)W^4_{m+n}+  \frac{N_3}{12}(m-n)(2m^2+2n^2-mn-8)L_{m+n}  
\nonumber \\
& & 
+\frac{8 N_3}{(c{\color{red} +\frac{22}{5}})}(m-n)\, \Lambda^{(4)}_{m+n} 
+ \frac{N_3 c}{144}m(m^2-1)(m^2-4)\delta_{m,-n}  \\
{}[W^3_m, W^4_n] &= & (3m-2n)W^5_{m+n} \nonumber \\
& & 
+ \frac{208 N_4}{25 N_3 (c{\color{red} + \frac{114}{7}}) }(3m-2n)\, \Lambda^{(5)}_{m+n}
+ \frac{84 N_4}{25 N_3 (c{\color{red} + 2})}\, \Theta^{(6)}_{m+n} 
\nonumber \\
& & 
- \frac{N_4}{15 N_3}(n^3 - 5m^3 - 3 m n^2 + 5 m^2 n - 9n + 17 m)W^3_{m+n} \ ,
\end{eqnarray}
where the $1/c$ corrections have been indicated in red. Similar corrections appear
at higher order, see \cite{CGKV}.

For the low-lying commutation relations given above, this is sufficient to solve the constraints 
coming from the Jacobi identities. However,  for the higher commutators we also get
conditions on the structure constants, i.e.\ on the analogues of $N_3$, $N_4$. In order
to describe this succinctly, it is convenient to rescale $W^3$ such that 
$N_3=\frac{2}{5}$, i.e.\ to choose $q^2 =  \frac{1}{8 (\mu^2-4)}$, and to redefine $W^4$ by 
\be\label{betadef}
\hat{W^4} = \beta^{-1} W^4 \qquad \hbox{with} \qquad 
\beta^2 = \frac{56}{75} \, \frac{N_4}{N_3^2} = \frac{4}{5} \, \frac{\mu^2-9}{\mu^2-4} \ .
\ee
(This redefinition has been chosen for convenience and the apparent singularities thus induced at $\mu^2=4$ in the expressions below are spurious and have no significance.)

As a result, the OPEs are of the form
\begin{eqnarray}\label{ww}
W^3\cdot W^3 & \sim & 
\frac{c}{3} \cdot {\bf 1} \ + \ 2 \cdot L \ + \ 8\, \sqrt{\frac{1}{5}\, \frac{\mu^2-9}{\mu^2-4}} \cdot 
\hat{W^4} + \cdots \\
W^3 \cdot \hat{W^4} & \sim & + \ 6\, \sqrt{\frac{1}{5}\, \frac{\mu^2-9}{\mu^2-4}} \cdot W^3 
+ \cdots  \ ,
\end{eqnarray}
and thus in the conventions of \cite{Hornfeck:1992tm}, the structure constant
$C^4_{33}$ satisfies
\be\label{c334c}
\left( C^4_{33} \right)^2  = \frac{64}{5} \, \frac{\mu^2-9}{\mu^2-4}
+ {\cal O}\left( \frac{1}{c} \right) \ . 
\ee
Note that we have included the possibility of an $1/c$ correction, given that we now know
that the algebra has to be corrected at that order. 

The Jacobi identities now imply that at least some of the higher structure constants are 
uniquely determined in terms of $C^{4}_{33}$ and $c$. For example, for the 
structure constants that  were calculated explicitly in 
\cite{Hornfeck:1992dp,Hornfeck:1993kp,Blumenhagen:1994wg,Hornfeck:1994is}, 
one finds \cite{Gaberdiel:2012ku} 
\begin{eqnarray}
C^4_{44} & = &  \frac{9(c+3)}{4(c+2)}\, \gamma - \frac{96 (c+10)}{(5c+22)}\, \gamma^{-1} \\[4pt]
(C_{34}^5)^2 & = & \frac{75 (c+7) (5c+22)}{16 (c+2) (7c+114)}\, \gamma^2 -25 \\[4pt]
C_{45}^5 & = &  \frac{15\, (17 c + 126) (c + 7)}{8\, (7 c + 114) (c + 2) } \, \gamma
-240 \frac{(c+10)}{(5c+22)}\, \gamma^{-1} \ ,
\end{eqnarray}
where 
\be
\gamma^2 \equiv \left( C^4_{33} \right)^2 \ . 
\ee
This suggests that at least these structure constants are fixed by the Jacobi
identities, and this was subsequently confirmed by an explicit analysis \cite{CGKV} where
in addition the next 40 or so structure constants were found to be determined uniquely 
in this manner. Note that there is a sign ambiguity in the definition of $C^4_{33}$, 
$C^5_{34}$, etc.; this is a consequence of the normalisation convention of 
\cite{Hornfeck:1992tm} which is defined by 
fixing the OPE of the spin $s$ field $W^{s}$ with itself
\be
W^{s} \cdot W^{s} \sim \frac{c}{s}\cdot  {\bf 1} + \cdots \ ,
\ee
and hence only determines the normalisation of each field up to a sign. We should also
stress that 
these relations modify the value of the structure constants in \w{\mu} relative
to those in ${\cal W}_{\infty}^{\rm cl}[\mu]$ by $1/c$ corrections; this justifies a posteriori
why we also included a $1/c$ correction in (\ref{c334c}).

Assuming that the Jacobi identities continue to determine all of these higher structure
constants, it then follows that the quantum \w{\mu} algebra is completely
characterised by the two parameters
\be\label{paras}
\gamma^2 \equiv \left( C^4_{33} \right)^2 \ , \qquad \hbox{and} \qquad
c \ .
\ee
Furthermore, we know that to leading order in $1/c$, the parameter $\gamma^2$ is determined 
by  the classical Poisson algebra ${\cal W}_{\infty}^{\rm cl}[\mu]$ to equal (\ref{c334c}),
i.e.\ $\gamma^2$ captures essentially the $\mu$-dependence of \w{\mu}. The fact that
we find a consistent 2-parameter family of \w{\mu} algebras characterised by (\ref{paras})
is therefore what one should have expected: it simply means that every classical 
${\cal W}_{\infty}^{\rm cl}[\mu]$ Poisson algebra can be quantised in a unique manner. 

The final step of the argument is to 
determine the {\em exact} $\mu$-dependence of $\gamma$; this can be 
done by employing the 
following trick. We know that, for $\mu=N$, \hs{N} can be truncated to $\mathfrak{sl}(N)$,
and we similarly expect that \w{N} can be truncated to ${\cal W}_N$. Thus the representation
theory of \w{N} must be compatible with the known representation theory of ${\cal W}_N$. 
Using this constraint, the {\em exact} $(c,\mu)$-dependence of $\gamma^2$ can
be determined \cite{Gaberdiel:2012ku} to be (see also 
\cite{Hornfeck:1992dp,Hornfeck:1993kp,Blumenhagen:1994wg} for earlier work
using essentially the same idea) 
\be \label{c433}
(C^4_{33})^2 \equiv \gamma^2 = 
\frac{64 (c+2) (\mu-3) \bigl( c(\mu+3) + 2 (4\mu+3)(\mu-1) \bigr)}{
(5c+22) (\mu-2)  \bigl( c(\mu+2) + (3\mu+2)(\mu-1)\bigr)} \ .
\ee
Note that (\ref{c433}) is indeed of the form (\ref{c334c}). The resulting algebra 
$\W_{\infty}[\mu]$ is now a well-defined $\W$-algebra for all values of $c$ and $\mu$.

\subsection{The Triality Relation}

The fact that \w{\mu} actually only depends on $\gamma^2$ (rather than directly on $\mu$)
has a very important
consequence. It means that the algebras \w{\mu} are equivalent for generically 
three different values of $\mu$. Indeed,  for given $c$ and $\gamma$, it
follows directly from (\ref{c433}) that the three values  are 
the roots of the cubic equation 
\be
\label{lambcub}
\bigl(3\tilde{\gamma}^2-8\bigr)\, \mu^3
+\bigl(\tilde{\gamma}^2(c-7)+(26-c)\bigr)\, \mu^2
-\bigl( 4\tilde{\gamma}^2(c-1)-9(c-2)\bigr)=0 \ ,
\ee
where we have defined $\tilde{\gamma}^2=\gamma^2\frac{(5c+22)}{64(c+2)}$.
Thus we have shown that 
\be\label{keyrelation}
\w{\mu_1} \cong \w{\mu_2} \cong \w{\mu_3} \qquad \hbox{at fixed $c$}
\ee
where $\mu_{1,2,3}$ are the roots of the cubic equation (\ref{lambcub}), evaluated
for a given $\gamma$. Note that the cubic equation does not have a linear term in 
$\mu$; thus the three solutions satisfy 
\be
\mu_1 \mu_2 + \mu_2 \mu_3 + \mu_3 \mu_1 = 0 \ ,
\ee
which is equivalent to $\sum_{i=1}^{3} \frac{1}{\mu_i}=0$ provided that
all $\mu_j\neq 0$. 

These algebras look very different from the point of view of $\hs{\mu}$ 
or even at the classical level. In fact, at very large $c$, eq.~(\ref{lambcub}) reduces to a 
linear equation in  $\mu^2$, and hence reduces to the familiar equivalence
between the classical $\w{\mu}$ algebras for $\pm \mu$ ---  this 
property is directly inherited from $\hs{\mu}$. The statement in (\ref{keyrelation}) is 
a very nontrivial generalisation to the quantum level (finite $c$), where 
the equivalence is a triality between the three values $\mu_{1,2,3}$. There
are three special cases where the cubic equation (\ref{lambcub}) 
degenerates: for $\mu=0$ we have 
$\tilde\gamma^2 = \frac{9 (c-2)}{4(c-1)}$, 
and the constant term in (\ref{lambcub}) vanishes. Then $\mu=0$ is a double
zero, and the other solution simply becomes 
\be
\w{\mu=0} \ \cong \ \w{\mu = c+1} \ .
\ee
For $\mu=1$, on the other hand, we have $\tilde\gamma^2 = \frac{8}{3}$, and the
cubic power vanishes; then we have the equivalences
\be
\w{\mu=1} \ \cong \ \w{\mu = -1}  \ \cong \ \w{\mu=\infty} \ .
\ee
The fact that for $\mu=1$ the symmetry $\mu\mapsto -\mu$ survives
at the quantum level is a direct consequence of the fact that, for this value of 
$\mu$, $\w{\mu}$ is a linear $\W$-algebra whose structure constants 
are simply the (analytic continuation of the) $\hs{\mu}$ structure constants.

Finally, the coefficient in front of the $\mu^2$ term in (\ref{lambcub}) vanishes
for $\tilde\gamma^2 = \frac{(c-26)}{(c-7)}$, when the equation becomes
$\mu^3=(c+1)$. Thus the three cubic roots of $(c+1)$ define equivalent 
$\w{\mu}$ algebras. 

\subsection{Triality in Minimal Model Holography}\label{sec:triality}

The above triality relation now allows us to prove that the asymptotic 
quantum symmetry of the higher spin gauge theory on AdS agrees exactly with the
$\W_{N,k}$ symmetry in the 't~Hooft limit. In order to see this, we take $\mu=N$,
and hence determine 
$\gamma =\gamma(\mu=N, c)$. Then it follows from (\ref{lambcub}) that
the other two roots $\mu_{2,3}$ satisfy the quadratic equation
\be\label{muceq}
\mu^2(N^2-1)-\mu(N-1-c)- N(N-1-c) =0\ ,
\ee
whose solutions are 
\be
\hspace*{-1cm} \mu_{2,3}(N, c) = \frac{1}{2(N^2-1)} \Bigl[
(N-1-c) \pm \sqrt{(N-1-c)(4 N^3 - 3N - c -1)} \Bigr] \ .
\ee
For the particular value $c=c_{N,k}$ defined in (\ref{cNk}), we then find
\be\label{Nkequiv}
\mu_2(N, c_{N,k})= \frac{N}{N+k} \qquad\hbox{and}
\qquad \mu_3(N, c_{N,k})= -\frac{N}{N+k+1}  \ , 
\ee
Thus we conclude, in particular, that the minimal model algebra ${\cal W}_{N,k}$
is isomorphic to 
\be\label{sident}
{\cal W}_{N,k} \cong {\cal W}_{\infty}[\lambda] \qquad \hbox{for} \ \lambda = \frac{N}{N+k} 
\quad \hbox{and at $c=c_{N,k}$.}
\ee
This therefore proves that the ${\cal W}$-algebra of the dual 2d CFT agrees indeed with
the quantisation of the classical symmetry algebra of higher spin gravity based on \hs{\lambda}. 
This correspondence is not at all obvious at the classical level, and is a very non-trivial
confirmation of the minimal model holography conjecture. We should also stress that (\ref{sident})
actually holds for {\em finite} $N,k$, not just in the 't~Hooft limit. This implies that the finite $N,k$
version of the duality should be constrained by this exact quantum symmetry.

We should also mention in passing that the other value of $\mu$, namely
$\mu_3= -\frac{N}{N+k+1}$, becomes in the large $N$ 't~Hooft limit $\mu_3=-\mu_2$.
This just recovers the by now familiar statement about the classical equivalence of the 
$\hs{\pm \mu}$ theories.  The relation between the different algebras can thus be summarised as

$$
\xymatrix@C=0.5pc@R=2.2pc{
 & & & & & \qquad \quad\; {\cal W}_{\infty}[\frac{N}{N+k}]  \; \ar@<-0.4ex>[llld] _{\cong}  \ar@<5ex>[dd]^{\cong}
 & \stackrel{\text{'t Hooft}}{\longrightarrow} & 
 {\cal W}_{\infty}[\lambda]\\
 \hbox{at $c=c_{N,k}$}   & {\cal W}_{N,k} \equiv {\cal W}_{\infty}[N]   & \ar@<0.4ex>[urrr] \ar@<-0.4ex>[drrr] _{\cong}
 & & & &  & (\lambda=\frac{N}{N+k})  \\
 & & & & & \quad \quad\;\; \;\;\; {\cal W}_{\infty}[-\frac{N}{N+k+1}]   \ar@<0,4ex>[lllu]   \ar@<-5ex>[uu]  
 & \stackrel{\text{'t Hooft}}{\longrightarrow} & 
 {\cal W}_{\infty}[-\lambda]\\
}
$$

\subsection{Relation to Coset Level-Rank Duality}

The above triality relation is in some sense an analytic
continuation of the conjectured level-rank duality of coset models 
\cite{Kuniba:1990zh,Altschuler:1990th}
\be\label{cosetdef}
{\cal W}_{N,k} \equiv
\frac{\mathfrak{su}(N)_k \oplus \mathfrak{su}(N)_1}{\mathfrak{su}(N)_{k+1}} \ \cong \ 
\frac{\mathfrak{su}(M)_l \oplus \mathfrak{su}(M)_1}{\mathfrak{su}(M)_{l+1}}   \equiv 
{\cal W}_{M,l} \ ,
\ee
where 
the relation between the parameters is 
\be\label{rel2}
k = \frac{N}{M} - N \ , \qquad l = \frac{M}{N} - M \ .
\ee
Here $M$ and $N$ are taken to be positive integers, whereas $k$ and $l$ are fractional 
(real) numbers, and the central charges of both sides are equal to 
\begin{eqnarray}\label{cNk1}
c_{N,k} & \equiv & (N-1) \, \Bigl[ 1 - \frac{N (N+1)}{(N+k) (N+k+1)} \Bigr]  \nonumber \\
& =  & 
(M-1) \Bigl[ 1 - \frac{M (M+1)}{(M+l) (M+l+1)} \Bigr]  \equiv  \, c_{M,l} \ .
\end{eqnarray}
If we assume that this level-rank duality will
also hold if instead of integer $N$, $M$, we consider the situation where $N$ and $k$
are integers, then we can solve (\ref{rel2}) for $M$ to obtain
\be
M \equiv \lambda = \frac{N}{N+k} \ ,
\ee
while $l$ is determined by the condition that both sides have the same central charge. 
Next we observe that we have also quite generically that
\be
\hspace*{-1cm}
\frac{\mathfrak{su}(M)_l \oplus \mathfrak{su}(M)_1}{\mathfrak{su}(M)_{l+1}}  \ \cong \ 
\hbox{Drinfel'd-Sokolov reduction of} \ \mathfrak{su}(M)\ \hbox{at level ${\hat{l}}$} \ , 
\ee
where again $\hat{l}$ is determined so as to have the same central charge as the left hand side.
For non-integer $M$ we can think of 
\be
\mathfrak{su}(\lambda) \cong \hs{\lambda} \ ,
\ee
and the Drinfel'd-Sokolov reduction of $\hs{\lambda}$ equals $\w{\lambda}$. Combining these statements
then leads to the claim that we have an isomorphism of algebras
\be\label{claim}
{\cal W}_{N,k} \equiv
\frac{\mathfrak{su}(N)_k \oplus \mathfrak{su}(N)_1}{\mathfrak{su}(N)_{k+1}} \ \cong \ 
\w{\lambda}   \qquad \hbox{with} \ \lambda = \frac{N}{N+k} \ .
\ee
Here the central charge of $\w{\lambda}$ is taken to agree with that of ${\cal W}_{N,k}$,
i.e.\ with $c_{N,k}$ defined in (\ref{cNk1}). This then reproduces (\ref{sident}).

Actually, there is a second variant of this relation. The ${\cal W}_N$ algebra at level $k$
is identical to the ${\cal W}_N$ algebra at level 
\be\label{khat}
k' = - 2 N - k - 1 
\ee
since the central charges of the two algebras agree, i.e. $c_{N,k} = c_{N,k'}$. 
Incidentally, this identification has a natural interpretation from the Drinfel'd-Sokolov (DS)
point of view. Recall that the cosets $\W_{N,k}$ in (\ref{cosetdef}) 
are equivalent to the DS reduction of $\mathfrak{su}(N)$ at level $\hat{k}\equiv k_{\rm DS}$, 
where 
the two levels are related as (see e.g.\ \cite{Bouwknegt:1992wg} for a review of these matters)
\be\label{DSrel}
\frac{1}{k+N} = \frac{1}{\hat{k}+N} - 1 \ .
\ee
{}From the DS point of view, replacing $k\mapsto k'$ as in (\ref{khat}) is equivalent
to replacing  $\hat{k}$ by $\hat{k}'$ with 
\be
\hat{k}' + N  = \frac{1}{\hat{k}+N}  \ .
\ee
In terms of the underlying free field description, this corresponds to exchanging
(see e.g.\ \cite{Bouwknegt:1992wg} or \cite[Section 6.2.2]{Gaberdiel:2011zw}) the roles of 
$\alpha_\pm$, i.e.\ to define 
$(\hat\alpha_+,\hat\alpha_-) = (-\alpha_-,-\alpha_+)$. 
This is an obvious symmetry of the DS reduction under which the representations
are related as $\Lambda_+\leftrightarrow \Lambda_-^\ast$. Thus we can repeat the 
above analysis with $k'$ in place of $k$, to conclude that 
$\W_{N,k'}$ is also equivalent to $\w{\mu}$ with $\mu=- \frac{N}{N+k+1}$; this then
reproduces also the third root $\mu_3$ in (\ref{Nkequiv}).

\section{Matching the Spectrum}

In the previous section we have shown that the symmetries of the higher spin theory
on AdS$_3$ and the proposed dual 2-dimensional CFT match in a rather intriguing
manner. Now we want to check that  the full spectrum of the two theories 
also agrees.  We only know how to calculate the spectrum of the higher spin theory
in the semi-classical regime, i.e.\ for $c\rightarrow \infty$; thus we can only
compare it to the CFT prediction in the 't~Hooft limit. 

We begin by studying the spectrum of the higher spin fields which, given the
results of the previous section, must agree with the vacuum representation of the CFT
in the 't~Hooft limit. From the 2d CFT point of view, modular invariance requires that 
the CFT also has other representations in its spectrum. By studying the 
finite $N$, $c\rightarrow \infty$ behaviour of these representations, we argue that 
some of them correspond to non-perturbative and some to perturbative states. We 
then explain that the contribution of the perturbative states are precisely reproduced 
by adding to the higher spin theory a complex massive scalar field. Finally, we review
a proposal for the interpretation of the remaining non-perturbative states as analytic continuations in $c$ of classical conical defect solutions.

\setcounter{footnote}{0}

\subsection{Higher Spin Fields}

The contribution of the massless higher spin fields to the 1-loop partition function
on thermal AdS$_3$ only requires a knowledge of their kinetic term. This can be most easily calculated using the Fronsdal description 
of higher spin fields \cite{Fronsdal:1978rb}. Taking carefully the various gauge
transformations into account, it was shown in \cite{Gaberdiel:2010ar} that the
contribution of a massless spin $s$ field to the 1-loop partition function equals
\be\label{dets}
\hspace*{-1cm}
Z_{(s)}^{\rm 1-loop}
=\Biggl[ \det\left(-\Delta  + \frac{s(s-3)}{\ell^2}  \right)^{\rm TT}_{(s)}\Biggr]^{-\frac{1}{2}}  \, 
\Biggl[ \det\left(-\Delta + \frac{s(s-1)}{\ell^2} \right)^{\rm TT}_{(s-1)}\Biggr]^{\frac{1}{2}}  \ ,
\ee
where `TT' means that only the transverse traceless part of the determinant is considered, 
and the index $(s)$ refers to the spin. (As before, $\ell$ is the AdS radius.)
Determinants of this form were explicitly evaluated in \cite{David:2009xg} using group
theoretic techniques; applying these results to the present context one finds that the 
$1$- loop answer factorises nicely into left and right moving pieces 
\be\label{detevl}
Z_{(s)}^{\rm 1-loop}= \prod_{n=s}^\infty {1\over |1-q^n|^2}\ ,
\ee
where $q=e^{i \tau}$ is the modular parameter of the boundary $T^2$ of the thermal 
background. This generalises the expression for the case of pure gravity $(s=2)$ 
\cite{Maloney:2007ud}, as explicitly checked in \cite{Giombi:2008vd}. 
Putting together the contributions of the fields of arbitrary spin $s=2,3,\ldots$, 
the total 1-loop contribution of the massless higher spin fields equals 
\begin{equation}\label{Zhs}
Z_{\rm hs}^{\rm 1-loop} =  \prod_{s=2}^{\infty} \prod_{n=s}^{\infty}\, \frac{1}{|1-q^n|^2} 
= |M(q)|^{2}\,  \times  \prod_{n=1}^{\infty} |1-q^n|^2 
\equiv |\tilde{M}(q)|^2 \ ,
\end{equation}
where $M(q)$ is the MacMahon function, and $\tilde{M}(q)$ is defined by 
\be\label{macm}
M(q)=\prod_{n=1}^{\infty} {1\over (1-q^n)^n} \ , \qquad
\tilde{M}(q) = \prod_{n=2}^{\infty} {1\over (1-q^n)^{n-1}}  \ .
\ee
The partition function $Z_{\rm hs}^{\rm 1-loop} $ in (\ref{Zhs}) now matches exactly the 1-loop 
contribution of the vacuum representation $|\chi_{(0;0)}(q)|^2$ 
of the ${\cal W}_{N,k}$ CFTs in the 't~Hooft limit. Indeed, 
by the usual Poincare-Birkhoff-Witt theorem (see for example \cite{Watts:1990pd}), 
a basis for the vacuum representation of ${\cal W}_\infty[\lambda]$ is given by
\begin{equation}\label{WNbas}
W^{s_1}_{-n^{1}_1} \cdots W^{s_1}_{-n_{l_1}^{1}} \, W^{s_2}_{-n^{2}_1} \cdots 
W^{s_2}_{-n^{2}_{l_{2}}} \cdots  
 W^{s_r}_{-n^{r}_1} \cdots W^{s_r}_{-n^{r}_{l_{r}}} \, \Omega \ ,
\end{equation}
where $s_1>s_2> \cdots > s_r\geq 2$ and 
\begin{equation}\label{basmod}
n^{j}_1\geq n_2^{j} \geq \cdots \geq n_{l_j}^{j} \geq s_j \ .
\end{equation}
Here we have used that $W^{s}_n\Omega = 0$ for $n\geq -s+1$ --- this is the reason for
the lower bound in (\ref{basmod}) --- but we have assumed that there are no other null vectors in
the vacuum representation, which is true in the 't~Hooft limit. (Note that we have 
denoted the Virasoro modes by $W^{2}_n\equiv L_n$.) Thus the character of the vacuum
representation equals
\be\label{chi00}
\chi_{(0;0)} = q^{-\frac{c}{24}} \, \prod_{s=2}^{\infty} \prod_{n=s}^{\infty} 
\frac{1}{(1-q^n)} \ .
\ee
The contribution of $|q|^{-c/12}$ in $|\chi_{(0;0)}(q)|^2$ corresponds to the tree level
part of the higher spin gravity calculation, and the remaining terms in (\ref{chi00}) 
then reproduce precisely (\ref{Zhs}).\footnote{A similar $1$-loop calculation in the parity 
violating topologically massive higher spin theory is suggestive of the vacuum character of a
logarithmic  $\W_N$ CFT \cite{Bagchi:2011td}.}

\subsection{Other States in the CFT}

As we have reviewed in Sec.~\ref{sec:minmod}, the minimal model CFTs 
also have other representations (apart from the vacuum representation).
As is familiar from rational CFTs, these representations
have to be present in the spectrum for a consistent (modular invariant) CFT.\footnote{Typically,
there will be more than one modular invariant combination of characters, and therefore more
than one consistent CFT. In the following we shall concentrate on the simplest
modular invariant, the `charge conjugation' theory, that exists for every rational CFT.} Note
that modular invariance is really a crucial ingredient in our analysis since the boundary of
thermal AdS$_3$ is in fact a torus, and hence  the possibility to go to finite temperature 
in AdS requires that the dual 2d CFT must be modular invariant (i.e.\ consistent on a
torus).

Recall that the most general representation of the ${\cal W}_{N,k}$ minimal model
is described by $(\Lambda_+;\Lambda_-)$, where $\Lambda_\pm$ are 
integrable highest weight representations of the affine algebra $\mathfrak{su}(N)$ at
level $k$ and level $k+1$, respectively. (Thus $\Lambda_\pm$ are Young diagrams
of at most $N$ rows, and at most $k$ and $k+1$ columns, respectively.) The simplest
representations (that generate all representations upon taking successive fusions)
are $({\rm f};0)$ and  $(0;{\rm f})$, as well as their conjugates, where ${\rm f}$
denotes the fundamental representation of $\mathfrak{su}(N)$. Their conformal
dimension equals (see eqs.~(\ref{h0f}) and (\ref{hf0})) 
\be\label{hf}
h({\rm f};0) = \frac{N-1}{2N} \Bigl( 1 + \frac{N+1}{N+k} \Bigr) \ , \qquad
h(0;{\rm f}) = \frac{N-1}{2N} \Bigl( 1 - \frac{N+1}{N+k+1} \Bigr) \ . 
\ee
In the 't~Hooft limit, they therefore become
\be\label{ftHooft}
\hbox{'t~Hooft limit:} \quad
h({\rm f};0) = \frac{1}{2} (1+\lambda) \ , \qquad
h(0;{\rm f}) = \frac{1}{2} (1-\lambda) \ .
\ee
However, in order to understand the nature of their duals in the hs theory, 
one should instead consider the limit where $N$ is being kept fixed, while 
$c\rightarrow \infty$ (the semi-classical limit)  \cite{Gaberdiel:2012ku}. In that limit, 
the two states behave rather differently, as one finds
\be\label{minlim}
\hbox{semi-classical:}\quad
h({\rm f};0) \sim - \frac{(N-1)}{2} \ , \qquad
h(0;{\rm f}) \sim - \frac{c}{2 N^2}   \ .
\ee
In particular, the conformal dimension of $(0;{\rm f})$ is proportional to $c$, thus
suggesting that this state should correspond to a non-perturbative (classical solution),
rather than to a perturbative excitation of the higher spin theory. Actually, a similar
consideration applies to any state for which $\Lambda_-$ is non-trivial. Thus one
is led to propose that only the states of the form $(\Lambda_+;0)$ should have
a perturbative origin in the higher spin theory \cite{Gaberdiel:2012ku}. We shall come back to 
the description of the remaining states (i.e.\ those with $\Lambda_-\neq 0$) in 
Sec.~\ref{sec:nonp}, but for the moment we now concentrate on these perturbative
states.

\subsection{Perturbative States}

It was proposed in \cite{Gaberdiel:2010pz,Gaberdiel:2012ku} that all CFT representations
of the form $(\Lambda_+;0)$ are accounted for by adding to the higher spin theory
a complex massive scalar of mass
\be\label{massc}
M^2 = - (1-\lambda^2) \ .
\ee
Recall from Sec.~2.1.4 that 
in the 3d higher spin theory of  \cite{Prokushkin:1998bq,Prokushkin:1998vn} (see also
\cite{Vasiliev:1999ba}), it is consistent to add a scalar multiplet to the higher spin theory,
but the mass of the scalar is then determined by the $\lambda$-parameter of the underlying 
\hs{\lambda} algebra as in (\ref{massc}). 

By the usual AdS/CFT dictionary, the mass of the scalar field is related to the conformal
dimension $\Delta$ of the corresponding conformal field; in 3d the relation takes the form
\be
M^2 = \Delta \, (\Delta - 2) \ .
\ee
Since $0\leq \lambda \leq 1$, $M^2$ in (\ref{massc}) lies in in the window $-1\leq M^2 \leq 0$, 
there are two real solutions for $\Delta$,  namely
\be
\Delta = (1 \pm  \lambda) \ . 
\ee
They correspond to the two different quantisations of the scalar field (since they characterise
two different asymptotic behaviours of the scalar field) \cite{Klebanov:1999tb}.
In the following we shall concentrate on the 
`usual' quantisation with $\Delta=1+\lambda$, for which $h=\bar{h}=\frac{1}{2}(1+\lambda)$. 
Note that this agrees precisely with the conformal dimension of the `fundamental' field
$({\rm f};0)$ or its conjugate, see eq.~(\ref{ftHooft}). 
\smallskip

The main evidence in favour of the above proposal comes from the comparison 
of partition functions \cite{Gaberdiel:2010pz,Gaberdiel:2011zw}. A real scalar field 
with boundary conformal dimension 
$h=\bar{h}=\frac{1}{2} \Delta$ contributes to the $1$-loop partition function on thermal AdS 
as \cite{Giombi:2008vd}
\be\label{scalar1intro}
Z_{\rm scal}^{\rm 1-loop}(h) = \prod_{j,j'=0}^\infty\frac{1}{1-q^{h+j}\bar{q}^{h+j'}} \ ,
\ee
and hence the contribution of a complex scalar is the square of (\ref{scalar1intro}). Note that
the form of (\ref{scalar1intro}) can be understood intuitively: a local operator of dimension 
$h$ has descendants which are obtained by acting on it with derivatives. Thus the 
`single particle' contribution to the  partition function is given by
\begin{equation}
Z_{\rm sing\, par}(h,q,\bar{q}) = {q^h \bar{q}^h \over (1 - q)(1 - \bar{q})}\ . 
\end{equation}
In the non-interacting limit, where we can neglect the anomalous dimensions of
composite operators, we can obtain the `multi-particle' partition function 
by using the standard formula for Bose statistics, leading to 
\be\label{scalar1}
Z_{\rm scal}^{\rm 1-loop}(h) = 
\exp{\left[\sum_{n=1}^{\infty} {Z_{\rm sing\, par}(h,q^n, \bar{q}^n) \over n}\right]} 
=  \prod_{j,j'=0}^\infty\frac{1}{1-q^{h+j}\bar{q}^{h+j'}} \ ,
\ee
thus reproducing (\ref{scalar1intro}).
For the comparison with the CFT calculation it is useful to rewrite $Z_{\rm scal}^{\rm 1-loop}(h)$
in terms of U$(\infty)$ characters following \cite{Gaberdiel:2011zw}. 
Recall that characters of $\un(N)$ in 
a representation $R$ are given by Schur polynomials in $N$ variables,
\be
\chi_R^{\un(N)}(z_i) = P_R(z_i) \ , \qquad i=1,\ldots, N \ .
\ee
Taking the large $N$ limit and evaluating on the Weyl vector, we can 
define the specialised Schur functions 
\begin{eqnarray}\label{schurpm}
P_R(q) &\equiv& \chi_R^{\mathfrak{u}(\infty)}(z_i) \ ,  \quad (z_i = q^{i - \frac{1}{2}}) \ , \\
P_R^+(q) &\equiv& q^{+\frac{\lambda}{2} B(R)} P_R(q) \ , \nonumber
\end{eqnarray}
where $B(R)$ is the number of boxes in the Young diagram $R$; explicit formulae for the
 Schur functions can be found in the appendix of \cite{Gaberdiel:2011zw}.
In terms of U$(\infty)$ characters, the scalar determinant (\ref{scalar1intro}) equals then
\be\label{scalpoly}
Z_{\rm scal}^{\rm 1-loop}(h) = \sum_{R} |P_R^{+}(q)|^2 \ .
\ee
Here the sum is over all Young diagrams of ${\rm U}(\infty)$, i.e.\ without any 
restrictions on the lengths of rows or columns. Combining the contribution
of two real (i.e.\ one complex) scalars then leads to 
\be\label{gravityfinal1}
Z_{\rm bulk}^{\rm pert} =  (q\bar{q})^{-c/24}\cdot |\tilde{M}(q)|^2\cdot 
\sum_{R,S}| P_{R}^+(q) P_{S}^+(q)  |^2 \ ,
\ee
where the sum runs over two sets of Young diagrams.

\subsubsection{Comparison to CFT}

This partition function should now be compared to the `perturbative part' of 
the CFT partition function, i.e.\ to 
\be\label{zcftschem}
Z^{\rm pert}_{\rm CFT}(N,k) = \sum_{\Lambda} |b_{(\Lambda;0)}(q)|^2 \ ,
\ee
where $\Lambda$ runs over all allowed representations of $\mathfrak{su}(N)_k$, and
$b_{(\Lambda;0)}$ is the branching function (i.e.\ the character) of the corresponding
${\cal W}_{N,k}$ representation, see eq.~(\ref{brfn}). 

\begin{figure}[htb]
\begin{center}
\epsfig{file=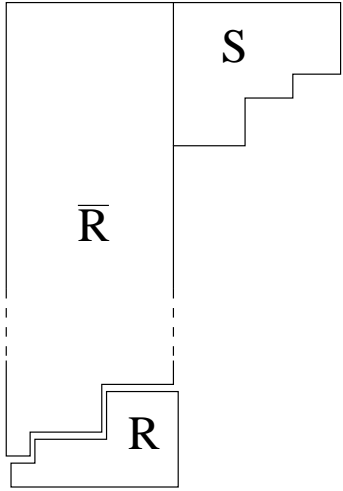}
\caption{\small A Young diagram of SU$(N)$ in the large $N$ limit.  
The full representation $\Lambda = (\overline{R}, S)$ has a finite number of `boxes'  
$S$ and `antiboxes' $R$.\label{fig:tableaux}}
\end{center}
\end{figure}

Since we can only calculate the gravity answer in the semi-classical limit, we need 
to take the $N\rightarrow \infty$ 't~Hooft limit, and hence have to be careful about 
which representations $\Lambda$ we should include.
As is familiar from similar situations, see e.g.\  \cite{Gross:1993hu}, a natural
prescription is to consider those representations 
$\Lambda$ that are contained in finite tensor powers
of the fundamental and anti-fundamental (where the number of tensor powers does not 
scale with $N$); note that the conformal dimension of $(\Lambda;0)$ 
is essentially proportional to the number of tensor powers in $\Lambda$, and 
hence this prescription takes account of all the low-lying representations of this type.
As in \cite{Gross:1993hu}, the corresponding Young diagrams can then be viewed as 
 two Young diagrams placed side by side,
\be
\Lambda = (\overline{R}, S) \ ,
\ee
where $\overline{R}$ is a tensor power of anti-fundamentals (`antiboxes') and 
$S$ is a tensor power of fundamentals (`boxes') as in Fig.~\ref{fig:tableaux}.  
We should also mention that the field identification (\ref{fieldidr}) becomes trivial in this 
limit since it does not lead to identifications among representations for which $R$ and $S$
are finite Young diagrams.

In order to calculate (\ref{zcftschem}) we next observe that the 
branching functions $b_{(\Lambda;0)}$ from Sec.~2.2.4, see eq.~(\ref{brfn}),
simplify considerably
in the 't~Hooft limit \cite{Gaberdiel:2011zw}. In particular, we can restrict the sum over the 
affine Weyl group to the finite Weyl group $W$, and we can simplify the exponent to arrive at 
\begin{equation}\label{brfnsimp}
b_{(\Lambda;0)}(q) \cong q^{\frac{(N-1)}{24} \lambda^2}\, 
 \frac{q^{C_2(\Lambda)+\frac{\lambda}{2} B} }{\eta(q)^{N-1}}  \, 
q^{\hat\rho^2}\,  
\sum_{w \in W} \epsilon(w) q^{-\langle w(\Lambda + \hat\rho), \,  \hat\rho\rangle }\ ,
\end{equation}
where $\cong$ denotes identities that are true up to terms that go to zero as $N\rightarrow \infty$,
and we have specialised to the case $\Lambda_-=0$ and written 
$\Lambda_+\equiv \Lambda = (\bar{R},S)$.
Furthermore, $B = B(R)+B(S)$  is the total number of boxes in the Young diagrams 
corresponding to $R$ and $S$. Following again  \cite{Gaberdiel:2011zw}, we can use 
the Weyl denominator formula for $\mathfrak{su}(N)$ 
\begin{equation}\label{weylde}
\sum_{w \in W} \epsilon(w) q^{-\langle w(\hat\rho), \hat\rho\rangle} = 
q^{-\hat\rho^2} \prod_{n=1}^{N-1} (1-q^n)^{N-n} \ ,
\end{equation}
which we solve for $q^{\hat\rho^2}$, to obtain
\be
\label{brfnsimp1}
b_{(\Lambda;0)}(q) \cong
q^{-\frac{c}{24}}\, 
q^{\frac{\lambda}{2} B + C_2(\Lambda) } \, 
\frac{\sum_{w \in W} \epsilon(w)\, q^{-\langle w(\Lambda +\hat\rho), \,  \hat\rho\rangle }}
{\sum_{w \in W} \epsilon(w) \, q^{-\langle w(\hat\rho), \hat\rho\rangle} }\, \tilde{M}(q) \ ,
\ee
where we have used that $c = (N-1) (1-\lambda^2)$ and $\tilde{M}(q)$ is as defined in 
(\ref{macm}). The ratio of sums in (\ref{brfnsimp1}) actually
equals the so-called quantum dimension of $\Lambda$, 
\be\label{qdim}
 \frac{S_{\Lambda 0}}{S_{00}}  = \dim_q(\Lambda) =
\frac{\sum_{w \in W} \epsilon(w)\, q^{-\langle w(\Lambda+\hat\rho), \, \hat\rho\rangle }}
{\sum_{w \in W} \epsilon(w) \, q^{-\langle w(\hat\rho), \hat\rho\rangle} } \ .
\ee
(Here $S_{ab}$ are the matrix elements of the $S$ modular transformation matrix
of the affine algebra.) 
Using results from \cite{Aganagic:2004js} and  \cite{Gross:1993hu}, one can show
that the quantum dimension of $\Lambda$ factorises as 
\be
q^{C_2(\Lambda)} \,  \dim_q(\Lambda) \cong q^{C_2(R)} \dim_q(R) \cdot q^{C_2(S)}\, \dim_q(S) \ ,
\ee
and for each finite Young diagram $L=R,S$ we have
\be\label{1.30}
\dim_q(L) = \chi_L^{\mathfrak{su}(N)} (\tilde{z}_i) =
\chi_L^{{\mathfrak u}(N)} (\tilde{z}_i)  = q^{-\frac{N}{2} B(L)} \, \chi_L^{{\mathfrak u}(N)} (z_i) \ , 
\ee
where  $B(L)$ is the number of boxes of $L$, and
\be
\tilde{z}_i = q^{-\frac{N}{2}}\, z_i \ , \qquad z_i = q^{i - \frac{1}{2}} \ .
\ee
Finally, using the large $N$ expansion of the quadratic Casimir (see 
\cite{Gaberdiel:2011zw} for details) it follows that 
\be
q^{C_2(L)} \, \dim_q(L) \cong \chi^{\mathfrak{u}(N)}_{L^T}(z_i) = P_{L^T}(q) \ , 
\ee
where $L^T$ is the representation whose Young diagram has been flipped relative to 
$L$, and we have used the notation introduced in eq.~(\ref{schurpm}). Inserting this
relation into (\ref{brfnsimp1}) we finally obtain
\be
\label{brfnsimp2}
b_{(\Lambda;0)}(q) \cong
q^{-\frac{c}{24}}\, 
P_{R^T}^+(q) \, P_{S^T}^+(q) \, \tilde{M}(q) \ .
\ee
Summing over all $R,S$ independently it is then obvious that $Z_{\rm CFT}^{\rm pert}$
in (\ref{zcftschem}) reproduces exactly $Z_{\rm bulk}^{\rm pert}$, see
eq.~(\ref{gravityfinal1}). This is a highly non-trivial check on the duality conjecture.

As an aside we should mention that in the original analysis of \cite{Gaberdiel:2011zw}, the 
above calculation was done both for the representations of the form $(\Lambda;0)$ and 
for those of the form
$(0;\Lambda)$. Furthermore, it was shown that the `light states', see Sec.~\ref{sec:nonp}
below,  decouple in the 't~Hooft limit, and therefore that the full CFT partition function (after 
removing the null-states that appear in the 't~Hooft limit) is exactly reproduced by adding
to the higher spin theory two complex scalar fields. However, this agreement only works
in the strict $N\rightarrow \infty$ limit; if we are interested in understanding the theory 
at finite $N$, we need to treat the states of the form $(0;\Lambda)$ differently.

\subsection{Non-Perturbative States}\label{sec:nonp}

As described in Sec.~5.2, only states in the CFT of the form $(\Lambda_+;0)$ have dimensions of 
order one in the semi-classical ($c\rightarrow \infty$, $N$ fixed) limit. 
Therefore we would like to  interpret all states $(\Lambda_+;\Lambda_-)$ with $\Lambda_- \neq 0$ 
as non-perturbative states in the bulk theory. 

To understand what these excitations might be, first focus on a class of states in the 
CFT of the form $(\Lambda_-;\Lambda_-)$, the so-called `light states'. The reason for this terminology
is that in the 't Hooft limit (as opposed to the semi-classical limit) these states are very light. 
Indeed, the dimension formula (\ref{hexpsimp}) gives 
\be
h(\Lambda_-; \Lambda_-)={1\over 2p(p+1)}\, 
\Bigl(\Lambda_- +2\hat\rho, \Lambda_- \Bigr) = \frac{C_2(\Lambda_-)}{(N+k)(N+k+1)}\ ,
\ee
which reduces in the 't~Hooft limit to (for $\Lambda$ having a finite number of boxes (S) or 
anti-boxes (R), in the notation explained below Fig.~1) 
\be\label{tHftlight}
\hbox{'t~Hooft limit:} \quad
h(\Lambda_-; \Lambda_-) = \frac{\lambda^2}{N^2}C_2(\Lambda_-) =
\frac{\lambda^2}{2N}(B(R)+B(S))\ .
\ee
Thus for finite $B(R),B(S)$, these dimensions go to zero and form a continuum of light 
states near the vacuum.  However, in the semi-classical limit they behave as 
\be\label{semilight}
\hbox{semi-classical:}\quad
h(\Lambda_-; \Lambda_-) \sim - \frac{c}{N(N^2-1)}C_2(\Lambda_-) + {\cal O}(1)\ ,
\ee
and thus are candidates for non-perturbative states. 
Here we have used the fact that 
\be\label{alpha0def}
\alpha_0^2 \equiv \frac{1}{(N+k)(N+k+1)} = \frac{(N-1-c)}{N(N^2-1)}  \rightarrow - \frac{c}{N(N^2-1)} \ .
\ee
In fact, in the semi-classical limit, it turns out that {\it all} states of the form $(\Lambda_+;\Lambda_-)$ have the same dimension i.e.
\be\label{semilight2}
\hbox{semi-classical:}\quad
h(\Lambda_+; \Lambda_-) \sim - \frac{c}{N(N^2-1)}C_2(\Lambda_-) + {\cal O}(1) \ ,
\ee
with only the ${\cal O}(1)$ terms depending on the representation $\Lambda_+$.

\subsubsection{Conical Defects}\label{sec:nonpert}

We will now outline how all these states $(\Lambda_+; \Lambda_-)$ (with $\Lambda_- \neq 0$)  
can be exactly accounted for, in the semi-classical limit 
(with $N$ fixed), by a class of solutions to the bulk equations of motion \cite{Castro:2011iw, Perlmutter:2012ds}. We first describe the solutions without scalar fields turned on. We can describe this sector by the Chern-Simons theory reviewed in 
Sec.~2.1. There are some important differences, which we will mention later, between the Lorentz 
signature theory, which has gauge group ${\rm SL}(N,\mathbb{R}) \times  {\rm SL}(N,\mathbb{R})$, 
and the Euclidean theory with gauge group ${\rm SL}(N,\mathbb{C})$. For the moment, we will 
consider the Lorentzian case and then mention the extension to the Euclidean setting later.

The equations of motion of the Chern-Simons theory are simply those for flat connections,
$F(A)=F(\bar{A})=0$. Therefore the only gauge invariant observables to characterise solutions 
are the holonomies of the gauge field. We will consider geometries which have the boundary 
topology of a torus. We will further look for solutions in which the topology is such that the spatial 
circle of the torus is contractible in the bulk while the time circle is not. This is therefore the same 
topology as global AdS$_3$. Note that for black holes, the role of the two circles is interchanged,
see \cite{AGKP}.

We now address the question of what the admissible (or smooth) classical solutions of the higher 
spin theory are.
The geometric notion of smoothness is somewhat subtle in a higher spin theory since the 
usual curvature invariants (which one uses to characterise smoothness) are actually not  
invariant under higher spin gauge transformations, see also \cite{AGKP} for a discussion
of this issue. However, in the present
case the higher spin gauge fields are simply ${\rm SL}(N)$ gauge fields,
and we can use our experience 
from gauge theory to rephrase the question. It is therefore natural to take the criterion to be that 
the gauge field configuration should not be singular. This is ensured if the  holonomy along a 
contractible curve is trivial (i.e.\ is gauge equivalent to the identity element). Otherwise the gauge 
connection would be singular somewhere in the interior of the curve. 

To see what this implies, let us fix a gauge and solve the equations of motion via the choice
(\ref{solc1}), (\ref{solc2}).  Then the holonomy
\begin{equation}\label{holo}
{\rm Hol}_\phi (A)={\cal P}\exp\Bigl(\oint_{S^1}  A\Bigr) = b^{-1}\exp(2 \pi a )\, b \ ,
\end{equation}
has to be  trivial, i.e.\ an element of the centre of the gauge group since the gauge fields are in the 
adjoint representation of the gauge group. This can be achieved if $\exp(2\pi a)$ is 
diagonalisable to an appropriate multiple of the identity matrix. 

\noindent
We can arrange this by choosing the $\mathfrak{sl}(N)$ gauge field to be of the form
\be
\label{condiag}
a =\sum_{ j=1}^{\lfloor N/2\rfloor} B^{(1)} _{2j-1}(n_j, n_j) \ ,
\ee
where the band $\mathfrak{sl}(N)$ matrices $B^{(1)}_k(a,b)$ are defined via
\be\label{bblock}
\Big[B^{(1)}_k(a,b)\Big]_{ij} = a \; \delta_{i,k}\delta_{j,k+1}- b \; \delta_{i,k+1}\delta_{j,k} \ .
\ee
Since $a$ in (\ref{condiag}) has eigenvalues $\pm i n_j$ ($ j=1 \ldots \lfloor N/2\rfloor$), the 
holonomy in (\ref{holo}) will be an element of the centre if we choose
\begin{eqnarray}
n_j  & \in  {\mathbb Z} \qquad  & {\rm for} \  a \in \mathfrak{sl}(N, {\mathbb R})  
\quad (N  \ {\rm odd}) \\[2pt]
n_j  & \in   {\mathbb Z} \quad \hbox{or} \quad n_j   \in   {\mathbb Z} + {\textstyle \frac{1}{2}}
\qquad  & {\rm for} \  a \in \mathfrak{sl}(N, {\mathbb R}) 
\quad  (N \ {\rm even})  \label{neven} \\[2pt]
n_j & \in {\mathbb Z} - {\textstyle \frac{m}{N}} \qquad &  {\rm for} \  a \in \mathfrak{sl}(N, {\mathbb C})\ .
\label{nquant}
\end{eqnarray}
This corresponds to the fact that the centre of ${\rm SL}(N,\mathbb{R})$ is ${\mathbb Z}_2$
for $N$ even while being trivial for $N$ odd. On the other hand for ${\rm SL}(N,\mathbb{C})$, the 
centre is ${\mathbb Z}_N$, and thus $m \in \{0,\ldots, N-1\}$ (independent of $j$). 

On the other hand, not all of these solutions satisfy the Brown \& Henneaux 
boundary conditions (\ref{asym}) which 
we needed for the asymptotic symmetry analysis. It can be shown \cite{Castro:2011iw} that the 
above solution can be brought to the highest weight gauge used in Sec.~3.2 if and only if 
the $n_j$ are all {\it distinct}.  

In the highest weight gauge we can easily read off the quantum numbers of the solution 
(mass, higher spin charges). Indeed, in that gauge the gauge field $a$ takes the form, 
generalising (\ref{ans3}) and the considerations that follow,  
\be\label{hwgauge}
a = L_1 + \sum_{s=2}^{N} a_s w^{(s)}_0 V_{-s+1}^{s} \ ,
\ee
where the $w^{(s)}_0$ are the spin $s$ charges, and $a_s$ is a suitable 
normalisation constant (see \cite{Castro:2011iw}). 
One can therefore express the charges $w^{(s)}_0$ in terms of traces of powers of $a$. Given the 
form of the solution (\ref{condiag}) with eigenvalues $\pm i n_j$ we have 
\be\label{Csdef}
{(- i)^s \over s} \tr (a)^s = {1 \over s} \sum_{i=1}^N (n_i)^s \equiv C_s (n)\ , \qquad 
s = 2,\ldots, N \ .
\ee
This then leads to \cite{Castro:2011iw}
\begin{eqnarray}\label{surplcharge}
w^{(2)}_0  &=&- \frac{c}{N(N^2-1)} C_2(n)\ , \nonumber \\
w^{(3)}_0 &=& - i {\Bigl(\frac{c}{N(N^2-1)}\Bigr)}^{3/2} C_3 (n) \ ,\\
w^{(4)}_0 &=& 
{\Bigl(\frac{c}{N(N^2-1)}\Bigr)}^2 \left( C_4 (n) - {C_4(\hat\rho) \over C_2(\hat\rho)^2 } 
C_2(n)^2\right)\ , \nonumber 
\end{eqnarray}
where $\hat\rho$ is the Weyl vector with components $\hat\rho_i=\frac{N+1}{2}-i$.
Note that in our conventions the vacuum AdS has $n_i=\hat\rho_i$ so that it has 
$w^{(2)}_0=L_0=-\frac{c}{24}$ and vanishing spin three and higher spin charges.

We can also write down the metric (in a particular gauge) from the explicit form of the 
gauge fields. For a large class of these solutions the metric is locally AdS with a conical 
surplus\footnote{We will loosely refer to the generic solutions as conical surpluses though not all of them can be viewed thus. 
One can also find a discrete spectrum of conical deficit metrics as solutions. They do not,
however, obey the boundary condition (\ref{asym}).}. We refer to \cite{Castro:2011iw} 
for more details. In \cite{Perlmutter:2012ds}, these solutions were further studied in the presence of a scalar field, leading to a rich spectrum of bound states of perturbative scalar quanta with the conical surpluses.  

\subsubsection{Comparison}

We can now compare this class of solutions with the non-perturbative states of the CFT. 
The key fact that we need is that 
\be
C_2(\Lambda)= {1\over 2}\sum_i  
\tilde n_i^2 - {\textstyle \frac{N(N^2-1)}{24}}  
= C_2(\tilde{n})  -   {\textstyle \frac{N(N^2-1)}{24} } \ , 
\ee
where the $\tilde{n}_i$ are distinct numbers given in terms of the row lengths $r_i$ of 
the corresponding Young diagrams
\be\label{tilden}
\tilde n_i = \Lambda_i + \hat\rho_i = r_i + {N+1 \over 2} - i - {B(\Lambda) \over  N}\ ,\label{ntildes}
\ee
and $B(\Lambda) =\sum_i r_i$ is the total number of boxes. We have also used the definition of 
$C_2(n)$ given in (\ref{Csdef}). With the identification $n_i=\tilde{n}_i$, the first line of 
(\ref{surplcharge}) agrees now,  up to the constant shift by the vacuum energy 
$\frac{c}{24}$, exactly with the spectrum of the states in (\ref{semilight2}), to leading order in $c$.
Note that both $n_i$ and $\tilde{n}_i$ are individually required to be distinct, 
and that  the $\tilde{n}_i$ in (\ref{tilden}) are indeed of the form (\ref{nquant}), which is the 
appropriate condition in Euclidean signature. 

One can similarly work out the higher spin charges of the $(\Lambda_+; \Lambda_-)$ 
states, at least in the 
semi-classical limit, and compare them to the other expressions in (\ref{surplcharge}).
As was shown in \cite{Castro:2011iw}, there is exact agreement in the 
semi-classical large $c$ limit (with fixed $N$). In \cite{Perlmutter:2012ds}, the comparison was carried further to include the ${\cal O}(1)$ terms and it was found that the pure conical surplus geometries have quantum numbers which are exactly those of the $(0; \Lambda_-)$ states (rather than the $(\Lambda_-, \Lambda_-)$ states as was originally proposed in \cite{Castro:2011iw}).

\subsubsection{Interpretation}

Given the above identification of the conical surpluses with the $(0; \Lambda_-)$ states, we can revisit the other states discussed in 
Sec.~5.2. As an illustration, consider the states $(\Lambda; {\rm f})$. We see from (\ref{minlim}) that  
$h(0;{\rm f}) \sim - \frac{c}{2 N^2} \sim h({\rm f};{\rm f})$ in 
the large $c$ limit. Indeed, we have, in this limit, 
\be\label{lambf}
h(\Lambda; {\rm f}) \sim h(0;{\rm f})  - (\Lambda, {\rm f}+\hat\rho) \sim h(0;{\rm f}) + h(\Lambda; 0)
- (\Lambda, {\rm f}) \ ,
\ee
where we have kept the term of order one but dropped terms of order $1/c$. Thus, as mentioned earlier, all 
states $(\Lambda; {\rm f})$  (with $\Lambda$ having a finite number of boxes, and anti-boxes) 
have approximately the same energy as the conical surplus $(0, {\rm f})$ in this limit. 
It can be shown that the sum of the last two terms in the r.h.s. of (\ref{lambf}) is always negative \cite{Perlmutter:2012ds}. Thus, the state $(0;{\rm f})$ is at the top of a band of states with energy spacings of order one. 
The proposal in \cite{Gaberdiel:2012ku} (modified suitably by \cite{Perlmutter:2012ds}, as we describe below) is to interpret all the non-perturbative states as bound states of the conical surplus with perturbative scalar excitations, 
i.e.\ the states $(\Lambda; 0)$.  
%In particular, we have the relation
%\be
%h(0; {\rm f}) = h({\rm f};{\rm f}) - h(\bar{\rm f};0) +  \frac{N-1}{N}  = h({\rm f}; {\rm f}) + \frac{(N-1)}{2} +  \frac{N-1}{N}  \ .
%\ee
%The first equality is exact, while the second only holds 
%in the semi-classical limit.  
%This is a prediction for an excited state of a conical surplus with a scalar field. 
Specifically, \cite{Perlmutter:2012ds}  proposes to identify the general state 
$(\Lambda_+; \Lambda_-)$ with $\Lambda_-\neq 0$ as a bound state of perturbative 
scalars $(\Lambda_+; 0)$ with a pure geometric surplus state 
$(0; \Lambda_-)$. Several pieces of evidence, including a matching of the quantum numbers to order one as well as the structure of null states, were provided in \cite{Perlmutter:2012ds}. 

Thus we now have fairly persuasive evidence for a candidate bulk dual for all states of the CFT, albeit in a semi-classical regime. 
This regime is related by analytic continuation in $c$ (keeping $N$ fixed) to the regime of the ${\cal W}_N$ minimal models
 $(c < (N-1))$. However, the primaries continue smoothly as we change $c$ and so we have evidence that the bulk 
 ${\rm hs}[\lambda]$ Vasiliev dual to the minimal models does capture all the states of the CFT.

\section{Further Checks}

In this section we briefly review a number of additional consistency checks that
have been performed: in Sec.~\ref{sec:corr} we discuss the matching of correlation
functions, while in Sec.~\ref{sec:BH} we explain the recent construction of black holes
and the calculation of their entropy.

\subsection{Correlation Functions}\label{sec:corr}

While the spectrum is an important check of the duality, more dynamical information is 
encoded in correlation functions. In particular, in a two dimensional CFT, the $3$-point 
function on the sphere is an important independent ingredient which then determines higher 
point functions via factorisation. We would like to match the CFT answer with the predictions 
from the bulk Vasiliev theory. Recall that this was the compelling piece of evidence 
\cite{Giombi:2009wh, Giombi:2010vg} for the Klebanov-Polyakov proposal for 
AdS$_4$/CFT$_3$ \cite{Klebanov:2002ja} and its generalisations 
\cite{Sezgin:2003pt}, see also \cite{Sundborg:2000wp,Witten,Mikhailov:2002bp,Sezgin:2002rt} 
for earlier work. Below we will review the calculations 
\cite{Chang:2011mz, Ahn:2011by, Ammon:2011ua, Chang:2011vk} that perform the 
analogous checks in the present case.

Another reason to study correlation functions has to do with the large $N$ limit. In gauge 
theories (or vector models), 't~Hooft's diagrammatic argument shows that the large 
$N$ limit is well defined (when we keep the 't~Hooft coupling fixed). In particular, if we 
normalise the $2$-point functions to be of order one, higher point functions of single trace 
operators are suppressed by inverse powers of $\sqrt{N}$. Furthermore, double trace operators 
behave like two particle states and thus their correlators can be factorised, to leading order in 
$N$, into those of the single particle states. 

While our coset CFTs seem to behave like a vector model, we do not have any general 
argument that the 't~Hooft limit defined in Sec.~1 leads to a familiar large $N$ expansion. 
For instance, the presence of a large number of light states (whose energy 
is proportional to $\frac{1}{N}$, see Sec.~5.4) could indicate 
that the $N\rightarrow \infty$ limit is not well behaved. In particular, even if 
every $3$-point function is suppressed by ${1\over N}$, this 
may not be sufficient to deduce a similar suppression for the $4$-point functions since
the large degeneracy of intermediate light states could potentially overcome the 
individual ${1\over N}$ suppression factors. It is therefore also 
important to check that the $4$-point functions are well behaved in the 
't~Hooft limit. We shall review below (see Sec.~6.1.2) the nontrivial checks 
on the factorisation of the $3$- and $4$-point functions 
that have been performed \cite{Papadodimas:2011pf, Chang:2011vk}.

\subsubsection{Three Point Functions}

The simplest class of $3$-point functions involve two scalar primaries with one 
higher spin current, $\langle {\cal O}\bar{\cal O}J^{(s)} \rangle$. Here 
${\cal O}$ denotes the scalar primary $({\rm f};0)$ which is dual to the 
perturbative scalar in the bulk (and $\bar{\cal {O}}$ is its complex conjugate).\footnote{The 
calculation can also be carried out analogously for the scalar primary 
$(0; {\rm f})$ which was later identified with a non-perturbative scalar
\cite{Gaberdiel:2012ku}.}
This correlator was first computed for small values of the 
spin $s$ and compared with the bulk calculation at 
$\lambda={1\over 2}$ in \cite{Chang:2011mz, Ahn:2011by}, and later generalised to 
arbitrary spin and $\lambda$ in \cite{Ammon:2011ua}; the answer is 
\be
\hspace*{-2cm}
\langle {\cal O}(z_1)\bar{\cal {O}}(z_2)J^{(s)} (z_3)\rangle 
= \frac{(-1)^{s-1}}{2\pi}\frac{\Gamma(s)^2}{\Gamma(2s-1)}
\frac{\Gamma(s+ \lambda)}{\Gamma(1+ \lambda)} 
\times \biggl(\frac{z_{12}}{z_{23}z_{13}}\biggr)^s\langle {\cal O}(z_1)\bar{\cal {O}}(z_2) \rangle \ .
\ee 
The CFT calculation in \cite{Ammon:2011ua}  assumes that the theory has 
$\w{\lambda}$ symmetry, and it follows from the triality described in Sec.~4, that this
is indeed the case for the 't~Hooft limit of the $\W_N$ theories.
On the bulk side, one uses the coupling of the scalar field to the higher spin 
gauge fields (\ref{Clin}) to compute the three point function, and finds exactly 
the same formula as the CFT answer from above. The computation makes clever use of the 
higher spin gauge symmetry to generate the solutions for the scalar field in the 
presence of the gauge fields. 

\subsubsection{Factorisation}  

The issue of large $N$ factorisation of correlation functions of the CFT was studied in 
\cite{Papadodimas:2011pf, Chang:2011vk}. Through explicit computation of a large 
number of correlators in the coset CFT using Coulomb gas and related techniques 
and then taking the large $N$ 't~Hooft limit, the following conclusions can be drawn:
\begin{itemize}
\item
Perturbative primaries built from multiple tensor powers of fundamental/anti-fundamental 
fields behave as multi particle states. Thus a primary such as $({\rm adj}; 0)$ behaves in 
$3$-point functions like a double trace operator --- the answer factorises, at leading order 
in $N$, into two $2$-point functions. 
\item
$4$-point functions of perturbative primaries also factorise at large $N$, and the light 
states do not appear in the intermediate channel at large $N$. They have a well 
defined large $N$ limit. 
\item
$4$-point functions of perturbative primaries such as $({\rm f}; 0)$ with non-perturbative 
primaries such as $(0; {\rm f})$ {\it also} factorise even though there are light states such as 
$({\rm f}; {\rm f})$ in the intermediate channel. The important point here is that the fusion rules 
of the CFT guarantee that of the very large number of light states only a finite number
propagates in the intermediate channel. Furthermore, the non-zero couplings are of 
order ${1\over N}$.  
\end{itemize}

Thus the perturbative primaries $(\Lambda ; 0)$ form a closed consistent subsector  
(at large $N$) for sphere amplitudes.  Furthermore, all of these states can be viewed 
as multi-particle states of a single complex scalar. Some of the non-perturbative states 
such as $(0; {\rm f})$ (and an infinite number of others at higher levels \cite{Chang:2011vk}) behave in much the same way as perturbative single particle states as far as their large $N$ behavior is concerned. 
Their correlation functions also have a well behaved 't~Hooft limit. However, because they 
essentially do not appear in any correlation function of perturbative states (unless 
there are order $N$ such operators), we can view them as a decoupled sector. As 
observed earlier, in the semi-classical limit these non-perturbative states indeed have 
$h \propto c$ justifying their name, even though in the 't~Hooft limit their dimensions 
are of order one.  

\subsubsection{Torus Two Point Function}
Let us also mention that in 
\cite{Chang:2011vk} the torus  $2$-point function of $({\rm f}; 0)$ and its conjugate
was calculated. This 
could potentially answer 
the question whether thermalisation occurs in these theories at large but finite $N$ 
at time scales small compared to the Poincare recurrence time which is $\sim N^4$\footnote{From the factorisation of correlators in the CFT, we know that we have a sum of terms like $q^{h+n}$, where $h$ are the conformal dimensions of various primaries and $n$ is an integer. From the form of $h$ given in eq.~(\ref{hco}), we see that it is a rational number with a denominator which goes like $N^4$ (the quadratic Casimir has a piece like ${1\over N}$). Therefore the Poincare recurrence time, i.e.\ the periodicity of the euclidean correlator in imaginary time, behaves as $N^4$.}. 
However, the explicit answer is not in a form which is easily amenable to a large $N$ 
expansion, and so more work needs to be done in order  to be able to extract interesting 
physics from it. A numerical study of the $N=2$ case does show encouraging signs of 
thermalisation occurring at intermediate time scales before recurrence sets in.

\subsection{Black Hole Entropy}\label{sec:BH}

As is implicit from the discussion in Sec.~\ref{sec:nonpert}, it is not immediately
obvious how to construct black hole solutions
in higher spin gravity. Indeed, the usual definition --- a spacetime singularity hidden 
behind a horizon --- is difficult to apply because neither the Riemann tensor nor 
the causal structure of the metric are gauge invariant. However, in Euclidean signature 
the problem is simpler, because a black hole is simply a smooth classical solution 
with torus boundary conditions. This definition has been used to construct explicit 
black hole solutions carrying higher spin charge \cite{Gutperle:2011kf}, see also 
\cite{Ammon:2011nk,Castro:2011fm,Tan:2011tj,Kraus:2011ds} as well as the
review \cite{AGKP} in this volume. 

The original construction of  \cite{Gutperle:2011kf} was done for spin $3$ gravity, but
this was later generalised to the case of the \hs{\lambda} higher spin theories in 
\cite{Kraus:2011ds}. The mass, angular momentum, and charges of the black hole were 
computed and used to infer the free energy \cite{Kraus:2011ds},
\begin{eqnarray}\label{gravres}
\log Z_{\rm BH}(\hat\tau, \alpha) & = &  \frac{i\pi c}{12\, \hat\tau} \Bigl[ 
1 - \frac{4}{3} \frac{\alpha^2}{\hat\tau^4} + \frac{400}{27} \frac{\lambda^2-7}{\lambda^2-4} \, 
\frac{\alpha^4}{\hat\tau^8} \nonumber \\ 
& & \qquad -\frac{1600}{27}\frac{5\lambda^4-85\lambda^2+377}{(\lambda^2-4)^2}
\frac{\alpha^6}{\hat{\tau}^{12}}+ \cdots \Bigr] \ ,
\end{eqnarray}
where $\alpha$ is the chemical potential for the spin-3 charge, and 
$\hat\tau$ is the complex structure of the torus, related to the black hole
temperature $T_H$ and (imaginary) angular potential $\Omega_H$ by
\be
\hat{\tau} = \frac{i}{2\pi T_H}(1 + \Omega_H) \ .
\ee
Furthermore, the central charge equals
$c=\frac{3\ell}{2G}$ with $\ell$ the AdS radius and $G$ Newton's constant. Note that 
(\ref{gravres}) only exhibits the holomorphic part of the full partition function; the
right-moving sector gives a similar contribution.

By the usual AdS/CFT dictionary, one expects (\ref{gravres}) to agree with the
CFT partition function 
\be\label{ZAdS}
Z_{\rm CFT}(\hat\tau, \alpha) = \Tr \Bigl( \hat{q}^{L_0 - \frac{c}{24}} \, y^{W_0} \Bigr)  \ , \qquad
\hat{q} = e^{2\pi i \hat{\tau}} \ , \quad y = e^{2\pi i \alpha} \ ,
\ee
in the high temperature regime,  i.e.\ for $\hat\tau\rightarrow 0$, and to leading order in the 
central charge $c$. Here $W_0$ is the zero mode of the spin $3$ current of 
${\cal W}_{\infty}[\lambda]$. Since (\ref{gravres}) is an expansion in powers of the 
chemical potential $\alpha$, it should be compared to the CFT expansion
\begin{eqnarray}\label{gravex}
Z_{\rm CFT}(\hat\tau, \alpha) & = &  \Tr \Bigl( \hat{q}^{L_0 - \frac{c}{24}} \Bigr) + 
\frac{(2\pi i \alpha)^2}{2!}  \Tr \Bigl( (W_0)^2\, \hat{q}^{L_0 - \frac{c}{24}} \Bigr)   \nonumber \\
& & + 
\frac{(2\pi i \alpha)^4}{4!}  \Tr \Bigl( (W_0)^4\, \hat{q}^{L_0 - \frac{c}{24}} \Bigr) + \cdots  \ .
\end{eqnarray}
At high temperatures, the $\hat{\tau}$-dependence of each term in the expansion is fixed 
by conformal invariance, which requires that \cite{Gaberdiel:2012yb}
\be\label{taudep}
\log Z_{\rm CFT}(\hat{\tau}, \alpha) \approx \frac{1}{\hat{\tau}}f\left(\alpha\over \hat{\tau}^2\right)
\ee
for some function $f$. As is familiar from entropy calculations \cite{Strominger:1997eq}, the 
standard method to obtain the partition function from
a dual conformal field theory point of view is to do the $S$-modular transformation
\be
\tau = -\frac{1}{\hat{\tau}} \ , \qquad q = e^{2\pi i \tau} \ .
\ee
In the high temperature regime, i.e.\ 
for $\hat\tau\rightarrow 0$,  we have $q\rightarrow 0$. The answer for the trace is then
dominated by the contribution from the vacuum state. 
This argument can be directly applied to the first term in the expansion 
(\ref{gravex}), 
\be\label{CFTex}
\Tr  \Bigl( \hat{q}^{L_0 - \frac{c}{24}} \Bigr) = \sum_{s,r} 
S_{s r}  \Tr_r \Bigl( q^{L_0-\frac{c}{24}} \Bigr) 
\sim \left(\sum_s S_{s 0}\right) \, q^{-\frac{c}{24}} + \cdots \ , 
\ee
where the sum runs over all primaries labelled by $r,s$ (with $r=0$ the 
vacuum representation), $S_{sr}$
is the modular $S$-matrix (not to be confused with the black hole entropy), and the dots 
indicate terms exponentially suppressed at high temperature. The
leading behaviour of the logarithm is then
\be
\log \Tr \Bigl( \hat{q}^{L_0 - \frac{c}{24}} \Bigr)  = - \frac{ i \pi c }{12} \tau + \cdots \ ,
\ee
and this reproduces precisely the $\alpha$-independent term in (\ref{gravres}),
using the relation $\tau=-\frac{1}{\hat\tau}$. This is equivalent to the Cardy formula for the 
entropy.  

In order to reproduce the subleading terms in (\ref{gravres}) from a CFT point of view
one therefore has to understand the modular behaviour of traces with the insertion of 
$W_0$ modes
\be\label{CFTtraces}
\Tr  \Bigl( \hat{q}^{L_0 - \frac{c}{24}}\, (W_0)^{2n} \Bigr)  
\ee
for $n=1,2,\ldots$ --- it is relatively easy to see that odd powers of $W_0$ will not
contibute at leading order in the high temperature expansion (\ref{taudep}). Using
the general transformation formula for torus correlation functions of 
conformal primary fields under modular transformations \cite{Zhu}, the leading
high temperature behaviour of (\ref{CFTtraces}) was determined for $n=1,2,3$ in 
\cite{Gaberdiel:2012yb}, thereby reproducing exactly (\ref{gravres}) from
(\ref{gravex}). As for (\ref{CFTex}), the calculation effectively only depends on 
the vacuum representation of the CFT, and hence does not probe the 
detailed spectrum of the conjectured dual. However, at least for $n=3$, various
non-linear terms of ${\cal W}_{\infty}[\lambda]$ contributed to leading order,
and hence the agreement is a pretty non-trivial test on the structure of 
${\cal W}_{\infty}[\lambda]$. The result is also in agreement with a direct
free field calculation  \cite{Kraus:2011ds} of (\ref{ZAdS}) that is available 
for $\lambda=0,1$, where we have a realisation of the CFT in terms of free fermions 
and free bosons, respectively.

The agreement between the two calculations demonstrates that the black hole solutions of
\cite{Gutperle:2011kf,Kraus:2011ds} dominate the bulk thermodynamics for
$T\rightarrow \infty$. However, it is currently not known whether Vasiliev gravity in 
three dimensions has a Hawking-Page transition, or whether the black hole dominates 
the bulk thermodynamics anywhere besides $T \rightarrow \infty$.  
If there is indeed a phase transition above which the black hole dominates, then the dual 
CFT should have a gap large enough so that (\ref{gravres}) applies above the transition 
temperature. The microscopic CFT proposed in \cite{Gaberdiel:2010pz} has a large 
number of light states with dimension $h+\bar{h} < 1$, so it presumably obeys 
(\ref{gravres}) only at asymptotically high temperatures. This is mirrored by the fact 
that the Vasiliev gravity theory  has other saddle point solutions \cite{Castro:2011iw}
(see Sec.~\ref{sec:nonpert}) which would contribute to the bulk thermodynamics.

\section{Generalisations}

In this section we sketch a number of relatively straightforward generalisations of the 
above duality conjecture.

\subsection{The Orthogonal Algebras}

The most obvious generalisation is the one that is analogous to the ${\rm O}(N)$ 
vector model in one dimension higher \cite{Ahn:2011pv,Gaberdiel:2011nt}:
it consists of replacing the ${\rm SU}(N)$ groups by ${\rm SO}(2N)$, 
i.e.\ it involves instead of (\ref{gencos}) the cosets
\be\label{socoset}
\frac{{\rm SO}(2N)_k \otimes {\rm SO}(2N)_1} { {\rm SO}(2N)_{k+1}} \ .
\ee
The ${\rm SO}(2N)$ groups have independent Casimir operators of even degree
$2,4,\ldots 2N-2$, as well as a Casimir operator of degree $N$, and thus 
the corresponding ${\cal W}$ algebra is generated by currents of the corresponding
spin $2,4,\ldots 2N-2$, as well as $N$. 
The algebra possesses a $\mathbb{Z}_2$-symmetry under which the spin-$N$ field is odd,
and the even subalgebra is then generated by the fields of even spin $2,4,\ldots, 2N-2$, together
with the normal ordered product of the spin $N$ field with itself and its higher derivatives, see \cite{Blumenhagen:1994wg}
for details. In the large-$N$ limit, we therefore
obtain a ${\cal W}$ algebra with
one current for every even spin. 

The central charge of the coset (\ref{socoset}) equals
\begin{equation}\label{central}
c = N \left[ 1 - \frac{(2N -1) (2N - 2)}{p (p+1)} \right] \ ,
\end{equation}
where  $p \equiv k+2N  -2$. The highest weight representations (hwr) of the coset are 
labelled by triplets $(\Lambda_+,\mu;\Lambda_-)$, where $\Lambda_+$ and 
$\Lambda_-$ are integrable hwr of $\mathfrak{so}(2N)_k$
and $\mathfrak{so}(2N)_{k + 1}$, respectively, while 
$\mu$ is a $\mathfrak{so}(2N)_{1}$ hwr. The triplets have
to satisfy the selection rule that $\Lambda_+ + \mu - \Lambda_-$ (interpreted as a 
weight of the finite dimensional Lie algebra $\mathfrak{so}(2N)$) lies in the root lattice of 
$\mathfrak{so}(2N)$. Modulo the root lattice, the weight lattice of $\mathfrak{so}(2N)$ has 
four conjugacy classes, and  there is precisely one level $1$ representation in each conjugacy 
class; thus the selection rule determines $\mu$ uniquely, and we can  label our 
coset representations by the pairs $(\Lambda_+;\Lambda_-)$. In addition there is the 
field identification $(\Lambda_+;\Lambda_-)\cong (A\Lambda_+;A\Lambda_-)$,
where $A$ is the outer automorphism of the affine algebra $\mathfrak{so}(2N)_k$ and 
$\mathfrak{so}(2N)_{k+1}$, respectively. $A$ permutes the four roots of the extended 
Dynkin diagram with Kac label $1$. As in the $\mathfrak{su}(N)$ case, the
field identification becomes irrelevant in the 't~Hooft limit.

We are again interested in the 't~Hooft limit, where we take $N$ and $k$ to 
infinity, keeping the ratio 
\begin{equation}
\lambda = \frac{2N}{k+2N-2} = \frac{2N}{p} 
\end{equation}
fixed. In this limit the conformal weight of the 
representations $(\Lambda;0)$ or $(0;\Lambda)$ that involve spinor labels is proportional to 
$N$, and the corresponding states decouple; for example, for the two spinor representations 
$s=[0^{N-2},1,0]$ and $c=[0^{N-1},1]$, one finds \cite{Gaberdiel:2011nt}
\begin{equation}
\hspace*{-1cm}
h_{(s;0)} = h_{(c;0)} = \frac{N}{8} \Bigl( 1 + \frac{2N-1}{p} \Bigr) \ , \qquad
h_{(0;s)} = h_{(0;c)} = \frac{N}{8} \Bigl( 1 - \frac{2N-1}{p+1} \Bigr) \ .
\end{equation}
Thus only the non-spinor representations survive. These are contained in 
tensor products of the vector representations and they have small
conformal dimension in the 't~Hooft limit; for example, for the vector representation 
$v=[1,0^{N-1}]$ itself we have
\begin{equation}\label{hvec}
\hspace*{-1cm}
h_{(v;0)} =   \frac{1}{2} \bigl( 1 + \frac{2N-1}{p} \bigr) \cong \frac{1}{2} (1 + \lambda) \ , \qquad
h_{(0;v)}  =   \frac{1}{2} \bigl( 1 - \frac{2N-1}{p+1} \bigr) \cong \frac{1}{2} (1 - \lambda) \ , 
\end{equation}
where we have denoted by $\cong$ the value in the 't~Hooft limit. The tensor products
of the vector representation can be labelled by Young diagrams, and thus the situation
is very similar to what was discussed above. There is only one small difference: 
the vector representation $(v;0)$ (and similarly for $(0;v)$) is its own conjugate representation, 
and thus there is no analogue of $(\bar{\rm f};0)$ in the current context.

Based on these observations one expects the dual higher spin theory to have
higher spin gauge fields of every even spin $s=2,4,6,\ldots$. In addition, one may guess
that the contribution of the representations that are contained in the tensor products of 
$(v;0)$ correspond to adding to the topological higher spin theory a {\em real} massive
scalar field of mass  \cite{Ahn:2011pv,Gaberdiel:2011nt}
\be
M^2 = - (1-\lambda^2) 
\ee
that is again quantised in the usual manner, i.e.\ leading to $h=\bar{h} = \frac{1}{2}(1+\lambda)$. 
It was shown in \cite{Gaberdiel:2011nt} that this proposal satisfies one important consistency
check: the spectrum of the higher spin theory together with this scalar field agrees 
exactly with the contribution of the perturbative $(\Lambda;0)$ states of the coset
(\ref{socoset}) in the large $N$ 't~Hooft limit. 

Unfortunately, the comparison of the partition functions does not directly
determine the underlying higher spin symmetry of the AdS theory (since the calculation
of the higher spin partition function only depends
on the quadratic part of the action). However, there is a proposal for what 
should replace \hs{\mu} in this context, namely the subalgebra
\be\label{hse}
\hs{\mu}^{(e)} \equiv \hbox{span} \{ V^s_m \in \hs{\mu} \; : \; s \ \hbox{even} \} \ .
\ee
In particular, the algebra $\hs{\mu}^{(e)}$ contains the `gravity' $\mathfrak{sl}(2)$ 
algebra generated by  $V^2_{0,\pm 1}$, and the Chern-Simons theory based on it will lead 
to spin fields of all even spacetime spins.
Recently, the quantum  ${\cal W}^{(e)}_{\infty}[\mu]$ algebra consisting of one conserved
current for every even spin was studied in some detail \cite{CGKV}. It was found that it is
again characterised in terms of two parameters, the central charge $c$ as well as 
the self-coupling constant of the spin $s=4$ field. The analogues of the 
triality relations of Sec.~4 were also derived, thereby proving the equivalence
of  the quantum symmetries. It was furthermore shown in \cite{CGKV} that the 
wedge algebra of ${\cal W}^{(e)}_{\infty}[\mu]$ becomes in the $c\rightarrow \infty$ limit
precisely ${\rm hs}[\mu]^{(e)}$, thereby proving that the higher spin theory
is indeed the one based on (\ref{hse}).

\subsection{The ${\cal N}=2$ Supersymmetric Models}

The bosonic higher spin theories we have discussed so far arise most naturally
from truncations of the ${\cal N}=2$ supersymmetric higher spin theories 
\cite{Prokushkin:1998bq, Prokushkin:1998vn}. These supersymmetric higher spin theories
have two (real) bosonic gauge fields of each spin $s=2,3,\ldots$, 
together with a single  current of spin $s=1$. In addition there are two (real) fermionic gauge 
fields for each spin $s=\frac{3}{2},\frac{5}{2},\ldots$. As in the bosonic case above, 
the structure of the theory depends on a real parameter $\mu$ that characterises
the underlying Lie algebra symmetry in the Chern-Simons formulation. For the 
supersymmetric case the relevant algebra is ${\rm shs}[\mu]$, which can be defined
in close analogy to \hs{\mu} in (\ref{Bdecomp}). To this end consider
\be
sB[\mu] = \frac{U(\mathfrak{osp}(1|2))}{\langle 
C^{\mathfrak{osp}} - \frac{1}{4} \mu (\mu-1) {\bf 1}\rangle} \ , 
\ee
where $\mathfrak{osp}(1|2)$ is the Lie algebra generated by $L_m$, $m=0,\pm1$ and
$G_r$, $r=\pm \frac{1}{2}$, with commutation relations
\begin{eqnarray}\label{N1}
{}[L_m, L_n] & = & (m-n) \, L_{m+n}  \nonumber \\
{}[L_m, G_r] & = & \Bigl(\frac{m}{2} - r\Bigr) \, G_{m+r} \\
\{G_r,G_s\} & = & 2 \, L_{r+s}  \ ,
\end{eqnarray}
and the Casimir operator $C^{\mathfrak{osp}}$ takes the form
\be\label{Csusy}
C^{\mathfrak{osp}} = C^{\rm bos} + {\textstyle \frac{1}{2}} C^{\rm fer} \equiv 
L_0^2 - {\textstyle \frac{1}{2}} (L_1 L_{-1} + L_{-1} L_1) 
+ {\textstyle \frac{1}{4}} \Bigl( G_{\frac{1}{2}} G_{-\frac{1}{2}} - G_{-\frac{1}{2}} G_{\frac{1}{2}} \Bigr) \ . 
\ee
By construction $sB[\mu]$ is an associative superalgebra with product $\star$, 
and we can make it into a Lie superalgebra by defining 
$[A,B]_{\pm} = A \star B \pm B \star A$. As before, the resulting Lie superalgebra contains
an abelian subalgebra generated by the identity ${\bf 1}$, and we define ${\rm shs}[\mu]$ by 
\be
sB[\mu] = {\rm shs}[\mu] \oplus \mathbb{C} \ ,
\ee
in close analogy to (\ref{Bdecomp}). By a straightforward calculation one shows that 
$C^{\rm fer}$, defined by (twice) the second term in (\ref{Csusy}), satisfies
\be
\Bigl( C^{\rm fer} \Bigr)^2 = C^{\rm bos} + C^{\rm fer} = C^{\mathfrak{osp}} 
+ {\textstyle\frac{1}{2} } C^{\rm fer} \ ,
\ee
and hence we can define orthogonal projection operators
\be
P_\pm = \frac{1}{2}\, \Bigl[  {\bf 1} \pm \frac{2}{(\mu-\frac{1}{2})} 
\bigl(C^{\rm fer}- {\textstyle \frac{1}{4}}\cdot {\bf 1}\bigr) \Bigr] \ ,\qquad
P_\pm^2 = P_{\pm} \ , \quad P_+ P_- = 0 
\ee
that commute with the bosonic subalgebra of ${\rm shs}[\lambda]$. Thus the 
bosonic subalgebra of ${\rm shs}[\lambda]$ actually decomposes as a direct sum into
\be
{\rm shs}[\mu]^{\rm bos} \cong \hs{\mu} \oplus \hs{1-\mu} \ ,
\ee
since on the image of $P_\pm$ the eigenvalue of $C^{\rm bos}$ equals
\be
C^{\rm bos} = C^{\mathfrak{osp}}  -  {\textstyle \frac{1}{2}} C^{\rm fer} = 
{\textstyle \frac{1}{4}\mu (\mu-1) 
- \frac{1}{2} \bigl\{ \mp \frac{1}{2} \bigl(\mu - \frac{1}{2} \bigr) + \frac{1}{4} \bigr\} } \ , 
\ee
i.e.\ either $C^{\rm bos} = \frac{1}{4}(\mu^2-1)$ or 
$C^{\rm bos}  = \frac{1}{4}(\mu^2 - 2 \mu) = \frac{1}{4}((1-\mu)^2-1)$. Finally, the analogue of 
(\ref{slN}) is now 
\be
{\rm shs}[\mu=-N]  / \chi_N \cong \mathfrak{sl}(N+1|N) \ . 
\ee

The above formulation is manifestly ${\cal N}=1$ supersymmetric --- (\ref{N1}) is the
wedge algebra of the ${\cal N}=1$ superconformal algebra --- but actually the theory
has ${\cal N}=2$ supersymmetry. In particular, the massless gauge fields organise themselves
into ${\cal N}=2$ multiplets as 
\be
{\textstyle (1\ \frac{3}{2} \ \frac{3}{2} \ 2) \qquad
(2\ \frac{5}{2} \ \frac{5}{2}\  3) \qquad (3\ \frac{7}{2} \ \frac{7}{2}\  4) \qquad \hbox{etc.}
} 
\ee
By analogy with the bosonic case, one expects that a massive scalar multiplet has to be 
added to the higher spin theory. In the supersymmetric case, each 
matter multiplet consists of a complex scalar field of mass 
\begin{equation}\label{eq:mass}
  M_\mu^2 = -1+\mu^2\ ,
\end{equation}
a Dirac fermion of mass $m_\mu$ with 
\begin{equation}
m^2_\mu= m^2_{1-\mu} = {\textstyle \big(\mu-\frac{1}{2}\big)^2} \ ,
\end{equation}
as well as a complex scalar and Dirac fermion of mass $M_{1-\mu}$ and
$m_{1-\mu}$, respectively. These fields must be quantised so that the 
corresponding conformal dimensions fit also into an ${\cal N}=2$ multiplet, i.e.\ as 
\be\label{susyscal}
\hspace*{-1cm}
{\textstyle \bigl[ \frac{1}{2} (1+\mu), \frac{1}{2} (1+\mu) \bigr]} \qquad 
\begin{array}{c}
{\textstyle \bigl[ \frac{1}{2} (1+\mu), \frac{\mu}{2}\bigr]} \\
{\textstyle \bigl[ \frac{\mu}{2},\frac{1}{2} (1+\mu)\bigr]} 
\end{array}
\qquad 
\begin{array}{c}
{\textstyle \bigl[ \frac{1}{2} (1+\mu), \frac{\mu}{2}\bigr]} \\
{\textstyle \bigl[ \frac{\mu}{2},\frac{1}{2} (1+\mu)\bigr]} 
\end{array} 
\qquad
{\textstyle \bigl[ \frac{\mu}{2}, \frac{\mu}{2}\bigr]} \ ,
\ee 
where the $[\frac{\mu}{2},\frac{\mu}{2}]$ field corresponds to the massive
scalar field with mass $M_{1-\mu}$, quantised in the non-standard fashion,
i.e.\ with $h=\bar{h} = \frac{1}{2} (1 - (1-\mu))=\frac{\mu}{2}$. 

\subsubsection{The Dual Kazama-Suzuki Models}

It was  proposed in \cite{Creutzig:2011fe} that the above higher spin theory 
with $\mu=\lambda$ is dual to the 't~Hooft like limit of a family of minimal ${\cal N}=2$ 
superconformal coset theories based on 
\begin{equation}\label{eq:cosets_manifest}
s\mathcal{W}_{N,k} = 
\frac{\mathfrak{su}(N+1)^{(1)}_{k+N+1}}{\mathfrak{su}(N)^{(1)}_{k+N+1}\oplus \mathfrak{u}(1)^{(1)}_{\kappa}} \ ,
\end{equation}
where $\kappa = N(N+1)(k+N+1)$ is the `level' of the $\un(1)$ algebra, and
the superscript `(1)' indicates that the relevant algebras are ${\cal N}=1$ 
supersymmetric.  (The ${\cal N}=1$ affine algebras are actually
isomorphic to a direct sum of the corresponding bosonic algebra (at a shifted level), together
with $\dim(\mathfrak{su}(N))$ free fermions.) The 't~Hooft limit consists again of taking 
$N,k$ to infinity, with $\lambda=\frac{N}{N+k+1}$ fixed. 

These cosets are manifestly $\mathcal{N}=1$ supersymmetric, but according to 
Kazama and Suzuki~\cite{Kazama:1988qp,Kazama:1988uz}, the actual chiral algebra
contains the ${\cal N}=2$ superconformal algebra. Geometrically, this is a consequence of 
the fact that the coset (\ref{eq:cosets_manifest}) is associated to the 
homogeneous space
\begin{equation}
  \mathbb{C}\mathbb{P}^N=\frac{{\rm U}(N+1)}{{\rm U}(N)\times {\rm U}(1)}\ ,
\end{equation}
which is actually a Hermitian symmetric space, i.e.\ possesses a complex structure. 
We should also mention in passing that (\ref{eq:cosets_manifest}) coincides with the 
Drinfel'd-Sokolov reduction of the affine superalgebra 
$\mathfrak{sl}(N+1|N)_{k_{\mathrm{DS}}}$ at level \cite{Ito:1990ac}
\begin{equation}
 k_{\mathrm{DS}}=-1+\frac{1}{k+N+1}\ . 
\end{equation}

Given that the ${\cal N}=1$ superconformal algebras are actually isomorphic to direct sums
of the corresponding bosonic subalgebras and free Majorana fermions, we can reformulate
the bosonic subalgebra of $\mathcal{W}_{N,k}$ in (\ref{eq:cosets_manifest}) as
\begin{equation}\label{eq:cosets}
s\mathcal{W}_{N,k}^{(0)} = 
 \frac{\mathfrak{su}(N+1)_k\oplus \mathfrak{so}(2N)_1}{\mathfrak{su}(N)_{k+1}\oplus 
 \mathfrak{u}(1)_{\kappa}}\ ,
\end{equation}
where $\mathfrak{so}(2N)_1$ is the bosonic algebra associated to the $2N$ free Majorana 
fermions
that survive after subtracting from the $N^2+2N$ free fermions of the numerator 
in (\ref{eq:cosets_manifest}) the $N^2$ free fermions of the denominator. The central 
charge of the coset algebra $s{\cal W}_{N,k}$ is therefore 
\begin{equation}\label{cdef}
c = (N-1) + \frac{k N (N+2)}{k+N+1} - \frac{(k+1) (N^2-1)}{k+N+1} = \frac{3kN}{k+N+1} \ .
\end{equation}
As reviewed in detail in \cite{Candu:2012jq}, the supersymmetric representations 
of the coset $s{\cal W}_{N,k}$ are labelled by $(\Lambda;\Xi,l)$, where 
$\Lambda$ and $\Xi$ denote hwr's of $\mathfrak{su}(N+1)_k$ and 
$\mathfrak{su}(N)_{k+1}$, respectively, while $l$ is an integer defined modulo $\kappa$.
The selection rule takes the form
\begin{equation}\label{eq:sel_rules_susy}
\frac{B(\Lambda)}{N+1}-\frac{B(\Xi)}{N}-\frac{l}{N(N+1)}\in \mathbb{Z}\ ,
\end{equation}
where $B(\Lambda)$ denotes the number of boxes in the Young diagram corresponding
to $\Lambda$, and similarly for $\Xi$; there are also field identifications (that are worked out in
\cite{Gepner:1989jq}), but they are again irrelevant in the 't~Hooft limit. The analogue of the
$({\rm f};0)$ representation of the bosonic theory is now the representation with 
$\Lambda={\rm f}$, $\Xi=0$, with $l=N$ because of (\ref{eq:sel_rules_susy}); its 
conformal dimension equals in the 't~Hooft limit, 
see e.g.\ eq.~(3.63) of \cite{Candu:2012jq}
\begin{eqnarray}
h({\rm f};0,N) & = &  \frac{N (N+2)}{2 (N+1) (N+k+1)} - \frac{N^2}{2 N (N+1)(N+k+1)} \nonumber \\
& = & \frac{N}{2(N+k+1)}  \cong  \frac{\lambda}{2} \ .
\end{eqnarray}
This reproduces the lowest conformal dimension of the scalar multiplet (\ref{susyscal}) with
$\mu=\lambda$. It was shown in \cite{Candu:2012jq} that the 1-loop partition function of 
the supersymmetric higher spin theory, together with the massive scalar multiplet   
(\ref{susyscal}), is reproduced exactly by the perturbative states (i.e.\ the states with $\Xi=0$)
of the above Kazama-Suzuki model in the 't~Hooft limit. It was also shown in 
\cite{Henneaux:2012ny,Hanaki:2012yf,Ahn:2012fz}
that the symmetries match at least partially, and the analogue of the quantum symmetry
analysis of Sec.~4 was recently performed in \cite{CG1}.  More recently, the smooth supersymmetric 
conical defect geometries in the bulk were studied in \cite{Tan:2012xi, Datta:2012km, Hikida:2012eu},
and it was suggested in \cite{Hikida:2012eu} that these classical solutions may account for all primaries of the 
dual CFT, as suggested by the analysis of \cite{CG1}.

\section{Questions and Future Directions} 

In the preceding sections we have outlined many of the features of the $\W_N$ minimal models 
and the evidence accumulated thus far, for a dual description, at large $N$, in terms of 
a classical higher spin theory on AdS$_3$. In the process, we have also exhibited the 
tractability as well as complexity of the CFT:
\begin{enumerate}
\item
The spectrum and partition function of the $\W_N$ minimal models 
are explicitly known for any $N$ (and $k$). Nevertheless, 
analysing the spectrum in the large $N$ 't~Hooft limit is quite subtle. We see the presence 
of a large number of light states $\Delta \sim {\cal O}({1\over N})$ --- a feature not seen thus 
far in other examples of the AdS/CFT correspondence.\footnote{See, though, 
\cite{Banerjee:2012gh} for a similar phenomena in 3d Chern-Simons vector models on 
$T^3$.} 
While we have concentrated on the states with $\Delta \sim {\cal O}(1)$ there is also a rich 
structure of primaries of dimension $N$ and higher which we have not touched upon.  
\item
Three and four point sphere correlation functions in the CFT can also
be explicitly calculated using conventional CFT techniques 
\cite{Papadodimas:2011pf, Chang:2011mz}. It is nontrivial that they have a sensible large 
$N$ limit which is consistent with a classical theory in the bulk. The two point torus 
correlator has also been computed for finite $N,k$ and clearly exhibits an intricate structure
 \cite{Chang:2011mz}. 
\end{enumerate}

The boundary theory therefore appears to be rich enough to serve as an insightful example of the 
AdS/CFT correspondence. In particular, unlike most studies of the AdS/CFT correspondence thus 
far, one may hope to use the CFT to learn about aspects of stringy/quantum gravity in AdS. 
Clearly, a first task is to build on existing studies of the spectrum and correlation functions to 
extract quantitative information about bulk physics. Specifically, one may envisage:
\begin{enumerate}
\item
{\bf Obtaining a more refined understanding of the spectrum of states from the bulk point of view.} 
We have identified the $(\Lambda, 0)$ primaries (with a finite number of boxes and anti-boxes) 
with perturbative multi-particle states of the complex scalar in the bulk. 
The $(\Lambda_+, \Lambda_-)$ primaries (with $\Lambda_- \neq 0$), on the other hand,
behave as nonperturbative states in the semi-classical (large $c$, finite $N$) limit, i.e.\  have
$ \Delta \propto c$. There is a class of non-trivial classical solutions in the bulk (the conical defects) 
whose quantum numbers match with those of the $(\star, \Lambda)$ primaries to leading order in 
$c$. It will be interesting to quantitatively check whether all the $(\Lambda_+, \Lambda_-)$ primaries
can indeed be viewed in the semi-classical Vasiliev theory as bound states of these defects
 with the perturbative scalar excitations\footnote{A precise proposal for this has recently been put forward in \cite{Perlmutter:2012ds} together with supporting evidence.}. 
\item
{\bf Understanding the significance of the light states in the bulk $\hs{\lambda}$ theory.} 
The identification of light states as conical defects is in the semi-classical ${\rm SL}(N)$ 
theory which is related by an analytic continuation in the central charge to the 
$\hs{\lambda}$ theory. Is there a way to understand these directly through some kind of 
quantisation of semi-classical solutions in the $\hs{\lambda}$ higher spin theory?
Can one give a more geometrical interpretation for them?
\item
{\bf Studying the interactions between perturbative and non-perturbative sectors.} 
The sector of non-perturbative primaries contains states which behave like single or 
multiparticle  excitations in correlation functions \cite{Papadodimas:2011pf, Chang:2011mz} 
with each other and with perturbative primaries. What is the meaning of this from the bulk?
\item
{\bf Understanding the primaries in the CFT whose dimensions grow like $N$ or higher. }
As mentioned before, the CFT also has primaries whose dimension grows at least
like $N$. Can these states be identified with micro states of black hole like solutions? Is there 
a phase transition 
at temperatures of order one where such states dominate the spectrum? 
Note that at asymptotically high temperatures (and thus very high energies) we have seen, in 
Sec.~6.2, a match of the states in the CFT with those of black holes in the bulk 
\cite{Kraus:2011ds, Gaberdiel:2012yb}. 
\item
{\bf Extracting thermal behavior from torus two point function.} We need to put the two point function computed in \cite{Chang:2011mz} into a form amenable to 
taking the large $N$ limit. The one may hope to see whether it exhibits exponential thermal decay for 
intermediate times much smaller than the Poincare recurrence time. This is related to the
previous question of whether we have black holes dominating the phase diagram at any 
finite temperature.  
\end{enumerate}
 
Symmetry is playing a very active role in this duality. Again, unlike other examples of 
AdS/CFT duality, here the matching of the global and gauge symmetries between the 
boundary and the bulk is a nontrivial dynamical fact. Specifically, from the bulk point of view, 
we have an $\hs{\lambda}$ classical gauge symmetry which is enhanced to the classical 
$\w{\lambda}$ asymptotic symmetry algebra. As we saw in Sec.~4, this is nontrivially equivalent 
to the large $N$ 't~Hooft of the $\W_N$ algebra of the boundary CFT. We believe this equivalence 
is pointing to directions worth exploring further:
\begin{enumerate}
\item
{\bf Quantum deformation of the bulk symmetry algebra.} At finite $N$ when we need to go to a 
quantum version of the Vasiliev bulk theory, the prediction is that the symmetry algebra is 
deformed to $\w{\lambda =\frac{N}{k+N}} \cong \W_N$. This requires a nonperturbative 
truncation of the symmetry currents to a maximal spin $s_{max}=N$.  This is reminiscent 
of the stringy exclusion principle that arises in other (stringy) AdS/CFT examples. 
\item
{\bf Integrability at the quantum level.} The above truncation immediately leads to the fact that instead of an infinite 
number of commuting conserved charges at the classical level, there are only finitely many at finite 
$N$. What does this mean for the integrability of the theory? Does it affect the physics of black holes 
in the theory? 
\item
{\bf Quantisation of the Vasiliev theory.} What kind of quantisation of the bulk can produce a
truncation like the above which would not be visible in the ${1\over N}$ expansion? Is there a 
naive first quantisation like those of strings which is adequate for the ${1\over N}$ expansion but 
not beyond? Is there a more geometric formulation of the quantisation in which the $\w{\lambda}$ 
symmetry plays a central role? The $\w{\lambda}$ algebra makes definite predictions for the exact
$c$-dependence of, for example, the mass of the scalar as well as the structure of the
higher spin algebra. Can one derive these corrections, 
at least to lowest order in $\frac{1}{c}$, directly from the higher spin theory point of view? 
\item
{\bf Proving the Duality.} Could the quantum $\w{\lambda}$ symmetry perhaps be powerful 
enough to prove the duality? One is looking for unitary representations of this algebra as well 
as modular invariance of the thermal partition function. Perhaps this constrains the matter 
primaries to be those of the $\W_N$ minimal models (up to the discrete choices of modular 
invariants). Alternatively, could one generalise the ideas in 
\cite{ Koch:2010cy,Douglas:2010rc,Jevicki:2011aa}  to the interacting CFTs considered here?
\end{enumerate}
 
We have discussed in Sec.~7 some of the generalisations of the original duality to 
orthogonal gauge groups as well as ${\cal N}=2$ supersymmetric cosets. There are many 
other avenues here as well:
\begin{enumerate}
\item
{\bf Other Modular Invariants.} Up to now we have focussed on the diagonal modular invariant 
while constructing the $\W_N$ CFT from its chiral sectors. There is a large class of other 
modular invariants as well which are also consistent CFTs, and it is natural to wonder whether 
large $N$ families of these admit higher spin AdS$_3$ duals.
\item
{\bf Massive deformations and RG flows.} The $\W_N$ minimal models have many relevant 
operators and it is possible to deform the CFTs by turning these on. Some of these RG flows, 
especially between nearby minimal models have been studied,  see e.g.\ \cite{Gaberdiel:2010pz},
being in some cases even integrable deformations.
It would be of obvious interest to have nice examples of holographic 
duals to such massive non-supersymmetric theories and their RG flows.      
\item
{\bf `Stringy Cosets'.} We can consider the general family of cosets
\begin{equation}\label{genlcos}
\frac{{{\rm SU}(N)}_k \otimes {\rm SU}(N)_l }{ {\rm SU}(N)_{k+l}} \ .
\end{equation}
If we define the 't~Hooft limit in this case with $k, l, N \rightarrow \infty$ with relative ratios held 
finite as in \cite{Kiritsis:2010xc}, then we find that the central charge grows like $N^2$. This is like 
in a gauge 
theory and it is natural to expect a stringy dual.\footnote{We thank Eric Perlmutter
for discussions about this idea.}
Indeed, the special case of $k=l=N$ recently 
studied in \cite{Gopakumar:2012gd} does arise as the low energy limit of a 2d gauge theory 
coupled to adjoint fermions. It would be very interesting to understand the string duals for 
these generically non-supersymmetric theories. These would also provide an embedding of the 
vector-like cosets into a larger string theory, perhaps along the lines of \cite{Chang:2012kt}.   
\item
{\bf de Sitter analogue.} Vasiliev higher spin theories can also be defined on dS spacetimes.
A dS$_4$/CFT$_3$ correspondence has been advanced for 4d Vasiliev theories 
\cite{Anninos:2011ui}. A similar attempt for the case of dS$_3$/CFT$_2$ seems to require 
an imaginary central charge for the CFT and other such undesirable features 
\cite{Ouyang:2011fs}. Are there, perhaps, ways around this?
\end{enumerate}  

We have not described the features of black holes and other classical bulk solutions in this theory, in any detail. There are tantalising hints here of a stringy generalisation of geometry and what it has to say about fundamental issues of singularities, existence of horizons etc.
Some of these issues will be addressed in the accompanying article in this issue 
\cite{AGKP}. 

To summarise, we expect various fruitful insights to emerge in the coming years from the 
study of minimal models and their holographic duals. 

\ack
We would like to thank our collaborators 
C.~Candu, A.~Castro, K.~Jin, M.~Gutperle, T.~Hartman, S.~Raju, J.~Raeymakers, 
A.~Saha, P.~Suchanek,  and C.~Vollenweider,  for sharing their insights and ideas. We 
would also like to thank S.~Gerigk, K.~Jin and C.~Vollenweider for their careful 
reading of the manuscript and comments. The work of MRG was partially supported
by the Swiss National Science Foundation, while the work of 
RG was supported in part 
by the Swarnajayanthi fellowship of the DST and more broadly by the generous 
funding of basic sciences by his fellow Indian citizens.

\end{document}